\documentclass[prodmode,acmjetc]{acmsmall} 

\usepackage[ruled]{algorithm2e}
\renewcommand{\algorithmcfname}{ALGORITHM}
\SetAlFnt{\small}
\SetAlCapFnt{\small}
\SetAlCapNameFnt{\small}
\SetAlCapHSkip{0pt}
\IncMargin{-\parindent}

\begin{document}

\markboth{M. Ahsan et al.}{Designing a Million-Qubit Quantum Computer Using Resource Performance Simulator}

\title{Designing a Million-Qubit Quantum Computer Using Resource Performance Simulator}
\author{Muhammad Ahsan
\affil{Duke University, USA}
Rodney Van Meter
\affil{Keio University, Japan}
Jungsang Kim
\affil{Duke University, USA}}


%
\begin{abstract} 
The optimal design of a fault-tolerant quantum computer involves finding an appropriate balance between the burden of large-scale integration of noisy components and the load of improving the reliability of hardware technology. This balance can be evaluated by quantitatively modeling the execution of quantum logic operations on a realistic quantum hardware containing limited computational resources. In this work, we report a complete performance simulation software tool capable of (1) searching the hardware design space by varying resource architecture and technology parameters, (2) synthesizing and scheduling fault-tolerant quantum algorithm within the hardware constraints, (3) quantifying the performance metrics such as the execution time and the failure probability of the algorithm, and (4) analyzing the breakdown of these metrics to highlight the performance bottlenecks and visualizing resource utilization to evaluate the adequacy of the chosen design. Using this tool we investigate a vast design space for implementing key building blocks of Shor's algorithm to factor a 1,024-bit number with a baseline budget of 1.5 million qubits. We show that a trapped-ion quantum computer designed with twice as many qubits and one-tenth of the baseline infidelity of the communication channel can factor a 2,048-bit integer in less than five months. 
\end{abstract}


\terms{Simulation tools, Quantum error correction and fault tolerance}


\acmformat{Muhammad Ahsan, Rodney Van Meter and Jungsang Kim, 2015. Designing a Million-Qubit Quantum Computer Using Resource Performance Simulator.}

\begin{bottomstuff}
This work was funded by Intelligence Advanced Research Projects Activity (IARPA) under the Multi-Qubit Coherent Operation (MQCO) Program and the Quantum Computer Science (QCS) program.

Author's addresses: Muhammad Ahsan, Department of Computer Science, Duke University; Rodney Van Meter, Faculty of Environment and Information Studies, Keio University; Jungsang Kim,Department of Electrical and Computer Engineering, Duke University.
\end{bottomstuff}

\maketitle

\section{Introduction}
Although quantum computers (QCs) can in principle solve important problems such as factoring a product of large prime numbers efficiently, the prospect of constructing a practical system is hampered by the need to build reliable systems out of faulty components~\cite{van-meter13:_blueprint}. Fault-tolerant procedures utilizing quantum error correcting codes (QECC) achieve adequate error performance by protecting the quantum information from noise, but comes at the expense of substantial resource investment~\cite{1}. The threshold theorem (or the quantum fault-tolerance theorem) says that a quantum computation of arbitrary size can be performed as long as the error probability of each operation is kept below a certain threshold value, and sufficient computational resources, such as the number of quantum bits (qubits), can be provided to implement adequate fault tolerance~\cite{Aharonov1997}. Although this is an encouraging theoretical result, an accurate estimate of the resource overhead remains an extremely complex task, as it depends on the details of the hardware (qubit connectivity, gate speeds and coherence time, etc.), the choice of protocols (QECC, etc.), and the nature of the target algorithms.
Several application-optimized architectures have been proposed and analyzed~\cite{13,vanMeterJETCS2008,21,MUSIQC,MonroeArXiv2012,GaliautdinovPRA2012,FowlerPRA2012},  yet the accurate quantification of resource-performance scaling for various benchmarks remains a challenging problem. 

In this work, we quantitatively define the {\em scalability} of a quantum architecture to mean that the resource overhead of running a quantum algorithm, while sustaining expected behavior in execution time and success probability (of order unity, $\sim O(1)$), increases linearly with the problem size. We propose a modular ion-trap-based architecture and quantify its scalability for three different benchmark circuits crucial for Shor's factoring algorithm~\cite{18}: a {\em quantum carry look-ahead adder} (QCLA)\cite{QCLA}, the CDKM {\em quantum ripple-carry adder} (QRCA)~\cite{QRCA}, and an {\em approximate quantum Fourier transform} (AQFT)~\cite{AQFT}. This architecture features  fast and reliable interconnects to ensure efficient access to computational resources, and enables flexible distribution of computational resources to various workload-intensive parts of the system depending on the circuit being executed. By evaluating this architecture for a variety of benchmarks, we show that it can achieve highly optimized performance by flexible and efficient utilization of given resources over a range of interesting quantum circuits.   

To quantify the performance of an architecture as a function of available resources, we develop a performance-simulation tool similar to those reported in Refs.~\cite{Map3}, \cite{whitney2007automated},\cite{QUALE}, and \cite{21} that (a) maps application circuits on to the quantum hardware, (b) generates and schedules the sequence of quantum logic gates from the algorithm operating on the qubits mapped to the hardware, and (c) estimates performance metrics such as total execution time and failure probability. Unlike  the tools reported previously, our tool features unique capabilities to (1) simulate performance over varying hardware device parameters, (2) allow dynamic resource allocation in the architecture, (3) provide detailed breakdown of resource and performance variables, and (4) enable visualization of resource utilization over (5) a range of benchmark applications. By leveraging these unique attributes we search the architecture space for a suitable QC design while providing valuable insights into the factors limiting performance in a large-scale QC.  


This paper is organized as follows: section~\ref{QCSec} describes benchmark quantum circuits and their characteristics. Section~\ref{ModSec} describes the underlying quantum hardware technology and the modular, reconfigurable architecture used in our simulation. The toolset and its main features are outlined in section~\ref{ToolSec}. Simulation results along with detailed discussions are given in section~\ref{SimSec}, and section~\ref{Extensibility} describes extensibility of the tool. Section~\ref{RelatedWork} puts our work in the context of other previous quantum architecture studies, and section~\ref{ConSec} summarizes the main insights gained from our study.

\section{Quantum Circuits} \label{QCSec}
\subsection{Universal Quantum Gates}
Quantum circuits consist of a sequence of gates on qubit operands. An $n$-qubit quantum gate perform a deterministic unitary transformation on $n$ operand qubits. In the terminology of computer architecture, a gate corresponds to an ``instruction'' and the specific sequences of gates translate into instruction-level dependencies. Similar to classical computers, it is known that an arbitrary quantum circuit can be constructed using a finite set of gates (called \emph{universal quantum gates}) which is not unique)~\cite{1}. For fault-tolerant quantum computation, one has to encode the qubits in a QECC, and perform logic gates on the encoded block of {\em logical qubits}~\cite{1}. There are two ways of performing gates on a logical qubit:  in the first procedure, the quantum gate on logical qubit(s) is translated into a bit-wise operation on constituent qubits (referred to as a ``transversal'' gate). Since an error on one constituent qubit in the logical qubit cannot lead to an error in another constituent qubit in the same logical qubit, the error remains correctable using the QECC and therefore a transversal gate is automatically fault-tolerant. For a good choice of QECC, most of the gates in the universal quantum gate set are transversal, and therefore fault-tolerant implementation is straightforward. Unfortunately, for most of the QECC explored to date (the class of QECC called additive codes), it is impossible to find a transversal implementation of \emph{all} the gates in the universal quantum gate set~\cite{ImpossTransv}. A second, general procedure for constructing such gates involves fault-tolerantly preparing a very special quantum state, called the ``magic state'', and then utilizing quantum teleportation to transfer the operand qubit into the magic state to complete the gate operation~\cite{ZhouPRA2000}. This operation is generally much more resource intensive and time consuming, so minimizing such operations is a crucial optimization process for the fault-tolerant circuit synthesis.

\begin{figure}[tb]
	\centering
	\includegraphics[width=0.8\textwidth]{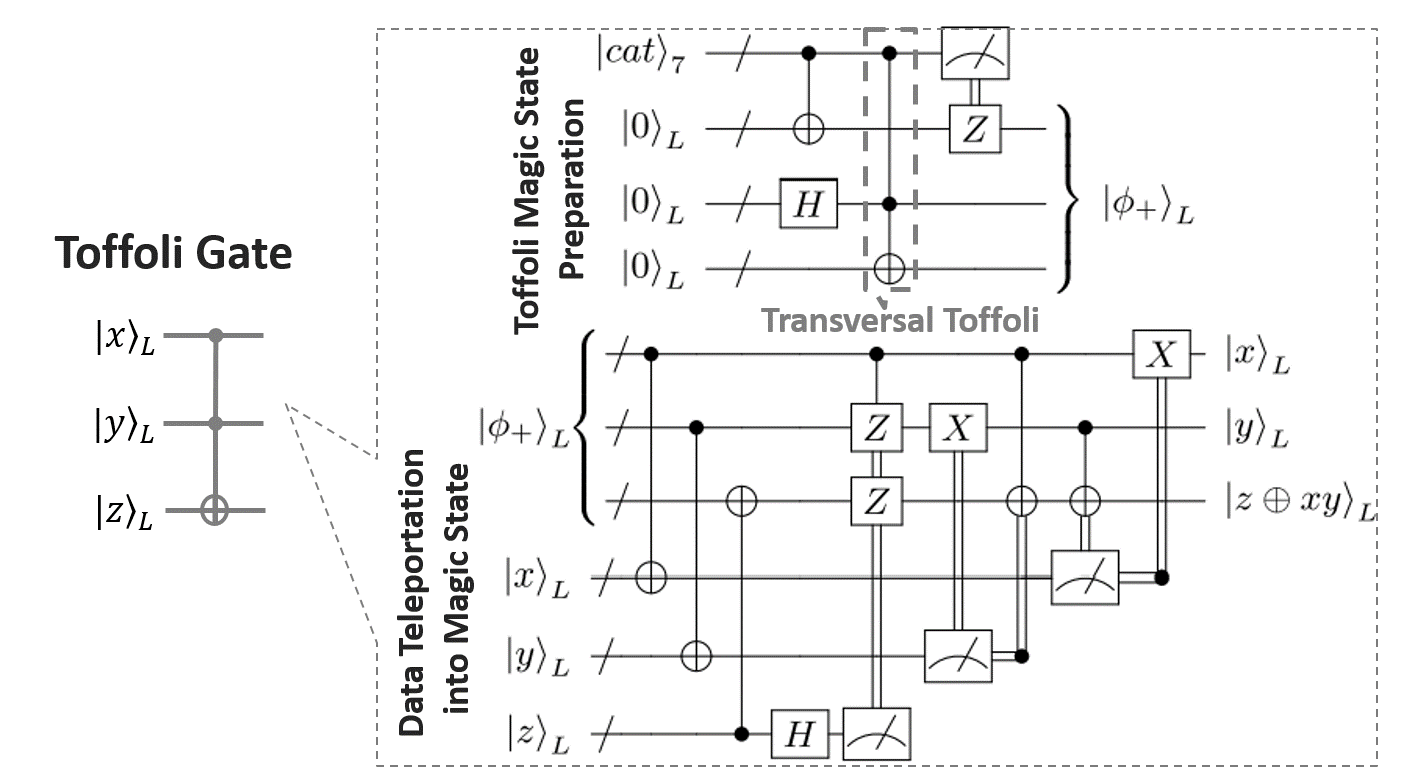}
	\caption{Fault tolerant circuit for Toffoli gate used in adder benchmarks. $|\alpha\rangle_{L}$ denotes a logical qubit block representing the state $|\alpha\rangle$.}
	\label{label:PicToff}
\end{figure}

We employ the widely used Steane [[7,1,3]] code~\cite{steane}. It invests seven qubits to encode one more strongly error-protected qubit. For the universal gate set, we utilize \{$X$, $Z$, $H$, CNOT, Toffoli\} for the adder circuits (both QCLA and QRCA), and \{$X$, $Z$, $H$, CNOT, $T$\} for the AQFT circuit.  For a single qubit state $\alpha |0\rangle + \beta |1\rangle$, $X$ and $Z$ correspond to the bit-flip and phase-flip operations that take the state to $\alpha|1\rangle+\beta|0\rangle$ and $\alpha|0\rangle - \beta|1\rangle$, respectively, and span a Pauli group of operators. $H$ is the Hadamard operator, which converts the computational basis states $|0\rangle$ and $|1\rangle$ to the equal linear superposition of two states $|\pm\rangle = (|0\rangle \pm |1\rangle)/\sqrt{2}$, and vice versa. CNOT is a two-qubit gate where the state of the second qubit (called the target qubit) is flipped {\em iff} the state of the first qubit (called the control qubit) is $|1\rangle$. Along with the Pauli operators, $H$ and CNOT span a Clifford group of operators (Clifford gates). In the Steane code, all operators in the Clifford group can be implemented transversally. However, in order to complete the set of universal quantum gates, a non-Clifford gate must be added. This could be either the $T$ gate (sometimes called the $\pi/8$-gate), a single-qubit gate which shifts the phase of the $|1\rangle$ state by $\pi/4$, or the Toffoli gate (sometimes called the Controlled-Controlled-NOT), a three-qubit gate where the state of the third qubit is flipped {\em iff} the state of the first and second qubits are both $|1\rangle$. Using either gate is equivalent, in the sense that a Toffoli gate can be constructed using several $T$ gates and Clifford gates~\cite{1}. Fault-tolerant implementation of a non-Clifford gate requires magic state preparation when the Steane code is used. Fig.\ref{label:PicToff} shows the fault-tolerant implementation of the Toffoli gate, where the Toffoli magic state is prepared with the help of four ancilla qubits followed by the teleporation of the operand qubits into the magic state.

\subsection{Benchmark Circuits}      
Shor's factoring algorithm consists of an arithmetic calculation called modular exponentiation~\cite{VanMeterPRA2005,vedral:quant-arith,beckman96:eff-net-quant-fact} which can be constructed from adder circuits, followed by a quantum Fourier transform~\cite{18}. Arithmetic circuits like adders (QCLA and QRCA) can easily be constructed from $X$, CNOT and Toffoli gates, while AQFT can be constructed more conveniently from $T$ gates. We consider a QC architecture where both Toffoli and $T$ gates can be executed, and optimize the architecture for executing all required quantum circuits for running Shor's algorithm.

\subsubsection{Quantum adders}
A large number of quantum adder circuits must be called to complete the modular exponentiation that constitutes the bulk of Shor's algorithm. We select two candidate adders QRCA and QCLA, representing two vastly different addition strategies, analogous to classical adders. QRCA is a linear-depth circuit, containing serially connected CNOT and Toffoli gates: an $n$-bit addition will require about $2n$ qubits to perform $2n$ Toffoli and $5n$ CNOT gates~\cite{QRCA}. The sequence of these gates is inherently local, and nearest-neighbor connectivity among the qubits is sufficient to implement this circuit. On the other hand, QCLA is a logarithmic-depth [$\sim 4\log_{2}n$] circuit connecting $4n$ qubits utilizing up to $n$ concurrently executable gates~\cite{QCLA}. This circuit roughly contains $5n-3\log_{2}n$ CNOT and Toffoli gates for $n$-bit addition. The exponential gain in performance (execution time) comes at the cost of sufficient availability of ancilla qubits and rapid communication channels among distant qubits to exploit parallelism. The QC hardware model considered here is unique in providing the global connectivity necessary for implementing QCLA. We study the resource-performance tradeoff in selecting QCLA vs. QRCA in Section~\ref{SimSec}.        

\subsubsection{Approximate quantum Fourier transform}
\begin{figure}[tb]
	\centering
	\includegraphics[width=0.8\textwidth]{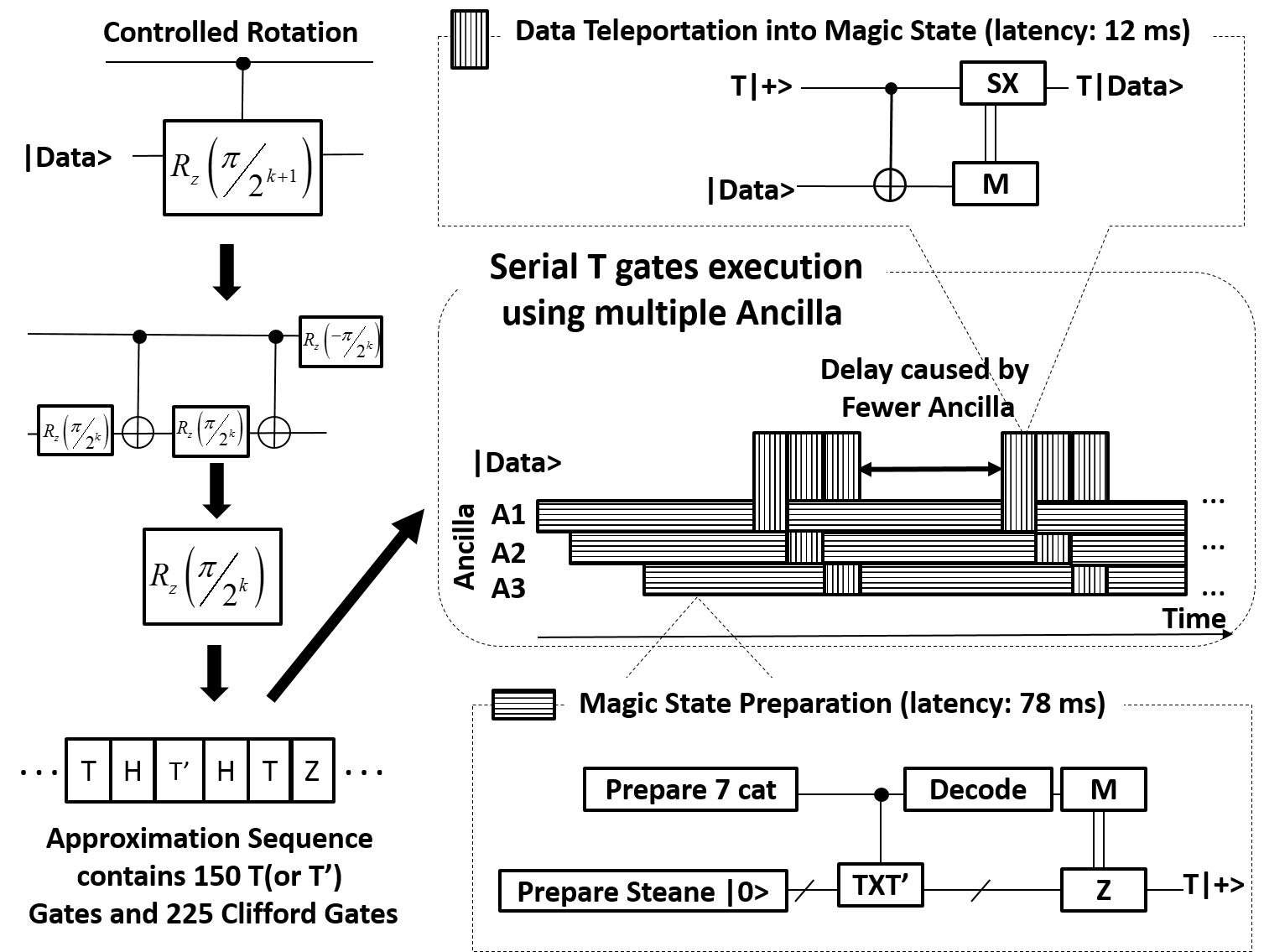}
	\caption{Fault-tolerant circuit for a controlled-rotation gate used in AQFT benchmark circuit. Small angle rotation gates are approximated by a sequence of $T$ and Clifford gates, and the $T$ gates are performed by magic state preparation and data teleportation.}
	\label{label:PicQFT}
\end{figure}

The quantum Fourier transform (QFT) circuit is often used as the keystone of the order-finding routine in Shor's algorithm~\cite{1}. It contains controlled-rotation gates $R_z(\pi/2^k)$, where the phase of the target qubit is shifted by $\pi/2^k$ for the $|1\rangle$ state if the control qubit is in the $|1\rangle$ state, for $1\le k \le n$, in a $n$-qubit Fourier transform. Fig.~\ref{label:PicQFT} shows that the controlled-rotation gates can first be decomposed into CNOTs and single-qubit rotations with twice the angle. These rotation operations are not in the Clifford group for $k > 1$, and must be approximated using gates from the universal quantum gate set~\cite{1}. A recent theoretical breakthrough provides an asymptotically optimal way of approximating an arbitrary quantum gate with a precision of $\epsilon$ using only $O(\log(1/\epsilon))$ Clifford group and $T$ gates, and a concrete algorithm for generating the approximation circuit~\cite{KliuchnikovPRL2013,GliesPRA2013}.

It has been shown that a QFT circuit can yield the correct result with high enough probability even if one eliminates all small-angle rotation gates with $k > 8$, sufficient to factor numbers as large as 4,096-bits \cite{AQFT}. The resulting truncated QFT is called the approximate QFT (AQFT). The depth of this benchmark circuit is linear in the size of the problem $n$, and the total number of controlled-rotation gates scales as $16n$. Using the method outlined in Ref.~\cite{KliuchnikovPRL2013}, we approximate rotations in our AQFT circuit with a sequence of 375 gates (containing 150 $T$ gates), with a precision of $10^{-16}$. The resulting approximation sequence consists of $T$ (or $T^\dagger$) gates sandwiched between one or two Clifford gates, whose execution time is negligible compared to the $T$ gate. The execution of the $T$ gate proceeds in two steps: preparation of the magic state $T|+\rangle$ and teleportation of data into the magic state. Since state preparation takes much longer time than teleportation in our system (78 ms vs 12 ms, see Table~\ref{label:T1}), we can employ multiple QC units to prepare magic states to simulate \emph{pipelined execution} of $T$ gates. When multiple ancilla qubits are available for the magic state preparation, we can reduce the delay in the execution of the approximation sequence. Using a simple calculation, we can show that the availability of 8 logical ancilla qubits  completely eliminates any delay. When the error correction procedure is inserted in the approximation sequence, its latency can be leveraged to eliminate the preparation delay with even fewer ancilla qubits.

\section{Quantum Hardware and Quantum Architecture Models}\label{ModSec}
\emph{Quantum hardware} describes the physical devices used to achieve computation using a specific technology~\cite{ladd10:_quantum_computers} such as trapped ions, atoms, superconductors, or quantum dots. The efficiency and reliability of QC depend on the characteristics of the chosen technology, such as execution time and fidelity of physical gate operations. We describe the physics of the quantum hardware by a set of device parameters (DP). In our simulation, the assumed baseline values for these parameters are optimistic, but can be  achieved in the near future through rapid technology advancement. Once quantum hardware technology is specified, we arrange the qubit resources according to their specific roles and their interconnection in order to assemble a large-scale QC. This is captured by the parameters of the \emph{quantum architecture}. For example, the number of qubits dedicated to perform fault tolerant quantum operations and the specification of communication channels are considered architecture parameters.

\subsection{Quantum Hardware Model} 
We choose to model quantum hardware based on trapped ions for its prominent properties that have been demonstrated experimentally. First, the qubit can be represented by two internal states of the atomic ion (\emph{e.g.}, $^{171}$Yb$^+$ ion~\cite{Yb-qubit}), described as a two-level spin system, manipulated by focusing adequate laser beams at the target ion(s). The physical ion qubits can be  individually accessible for computation~\cite{KnoernschildAPL2010,CrainAPL2014}. These qubits can be reliably initialized to the desired computational state and measured with very high accuracy using standard techniques. Most importantly, by virtue of the very long coherence time of the ions, qubits can retain their state (memory) for a period of time unparalleled by any other quantum technology. The qubit memory error is modeled as an exponential decay in its fidelity $F \sim \exp(-a t)$, where $a$ (=$1/T_{coh}$) is determined by the coherence time of the qubit, and $t$ is the time between quantum gates over which qubit sits idle (\emph{no-op}). The corruption of the qubit state is modeled using a depolarizing channel~\cite{1} (equal probability of bit flip, phase flip and bit-and-phase flip errors). Arbitrary single qubit gates, CNOT, and measurement can be performed with adequate reliability, making trapped ions a suitable candidate for large scale universal QC. 

A single-qubit quantum gate is accomplished by a simple application of laser pulse(s) on the qubit in its original location. A two-qubit gate, on the other hand, requires that both ions are brought in proximity before the laser pulse(s) are applied.  In our model, there are two ways to achieve this proximity using two different types of physical resources: the \emph{ballistic shuttling channel} (BSC) and the \emph{entanglement link} (EL)~\cite{MonroeArXiv2012}. BSC provides a physical channel through which an ion can be physically transported from its original location to the target location by carefully controlling the voltages of the electrodes on the ion trap chip. This chip can be modeled as a 2-D grid of ion-trap cells as shown in Fig.\ref{label:Pic0}. The dimensions of the state of the art ion-trap cell described in \cite{MonroeScience2013} fall in the $\sim$mm size range, and we use $T_{shutt} = 1\mu s$ as the time it takes for an ion to be shuttled through a single cell.
In the EL case, an entangled qubit pair (also known as the Einstein-Podolski-Rosen, or EPR, pair) is established between designated proxy ``entangling ions'' (e-ions) that belong to two independent ion trap chips using a photonic channel. This process is called \emph{heralded entanglement generation}~\cite{DuanQIC}, since the successful EPR pair generation is announced by the desired output of detectors collecting ion-emitted photons. The resulting EPR pair is used by the actual operand ions as a resource to perform the desired gate via quantum teleportation between two ions that cannot be connected by BSC~\cite{GottesmanNature1999}. It should be noted that the generation time for the EPR pairs is currently a slow process due to technology limitations. This slowness can be compensated for by generating several EPR pairs in parallel using dedicated qubits and hardware. Table~\ref{label:T0} summarizes the DPs used for all the analyses in this paper.

	\begin{table} \tbl{Device Parameters (DPs)\label{label:T0}}{
			\begin{tabular}[c]{ |c|c|c| }	
				\hline
				Physical Operation & Time($\mu s$) & Failure Probability\\ \hline \hline 
				Single-qubit  & $1$ & $10^{-7}$    \\ \hline
				Two-qubit (CNOT)  & $10$ & $10^{-7}$   \\ \hline
				Three-qubit (Toffoli) & $100$ & $10^{-7}$  \\ \hline
				Measurement  & $100$ & $10^{-7}$   \\ \hline
				EPR pair Generation  & $5000$ & $10^{-4}$    \\ \hline 
			\end{tabular}}
	\end{table}

\subsection{Quantum Architecture Model}
\label{subsec:Architecture}

\begin{figure}[tb]
	\centering
	\includegraphics[width=0.8\textwidth]{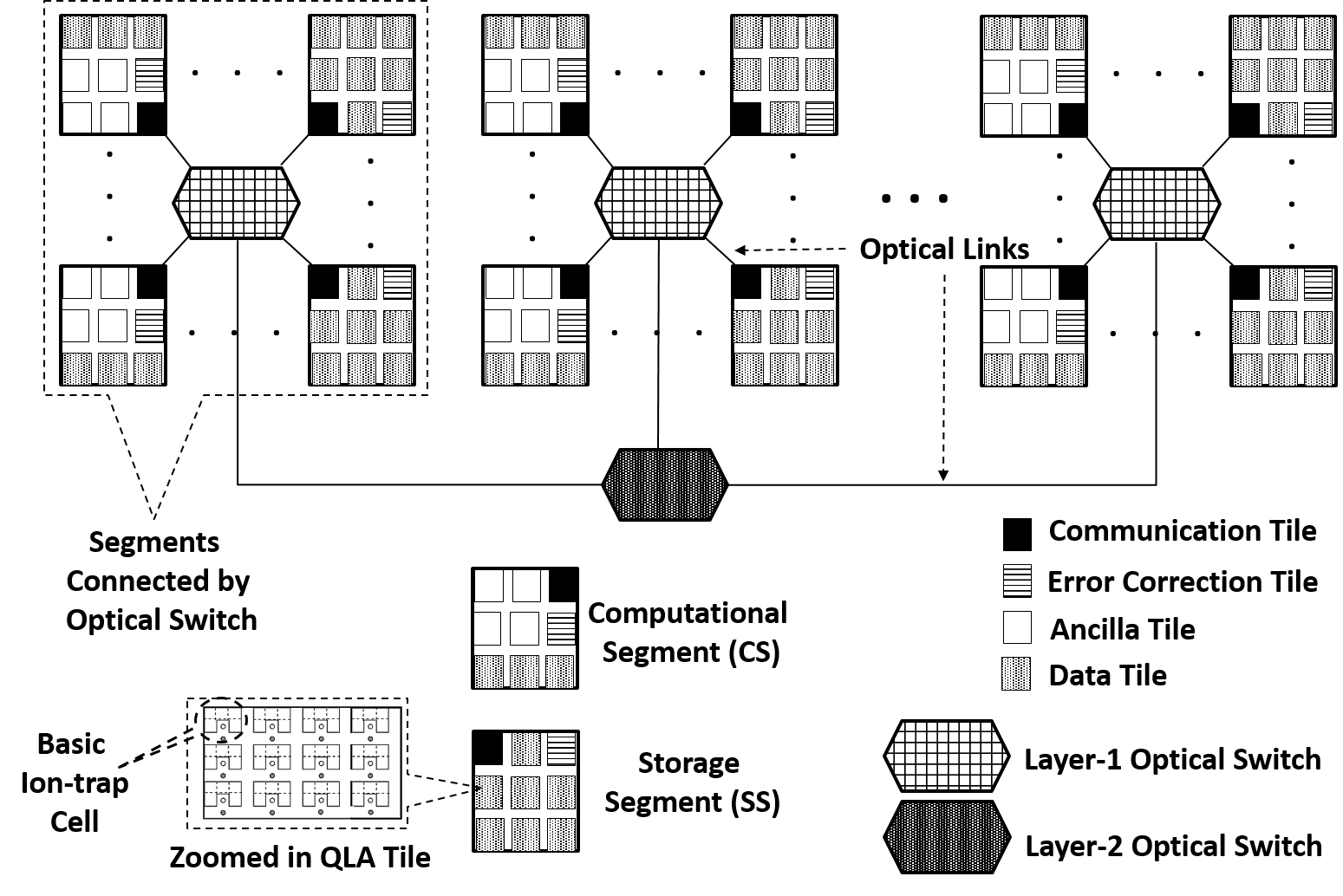}
	\caption{Overview of the reconfigurable quantum computer architecture analyzed in our performance simulation tool.}
	\label{label:Pic0}
\end{figure}

Our model is similar to the modular universal scalable ion-trap QC (MUSIQC) architecture~\cite{MonroeArXiv2012} shown in Fig.\ref{label:Pic0}. It features a hierarchical construction of larger blocks of qubits (called {\em segments}) composed of smaller units (called {\em Tiles}), which are connected by an optical switch network. We use the Steane [[7,1,3]] code~\cite{steane} to encode one logical qubit using 7 physical qubits. Additional (ancilla) qubits are supplied to perform error correction and fault tolerant operations on the logical qubit. We first construct the first layer (L1) logical qubit block containing 22 physical qubits (7 data and 15 ancilla qubits) using Steane encoding. As the size of computation grows, multiple layers of encoding are needed to minimize the impact of increasing noise: with each new layer, the qubit and gate count increase by about a factor of 7. At least two layers of encoding are essential for reliable execution of the sizable benchmark circuits analyzed in our simulations. Therefore, we cluster 7 L1 blocks to construct the second layer (L2) logical qubit block containing dedicated qubits to simultaneously carry out error correction operations at L1 level after every L1 gate. We find that the error correction operation at L2 level occurs much less frequently, and a dedicated error correcting ancilla resource at L2 is not necessary for each L2 logical qubit. Therefore, we allocate fewer ancilla qubits at L2 level for error correction, and rely on resource sharing to accomplish fault-tolerance at L2 level. 

At L2 level, we construct four different types of logical qubit blocks using L1 logical blocks, called {\em L2 Tiles}, that serve different functions in the computation~\cite{13}. Each Tile consists of memory cells that provide storage and manipulation of qubits for quantum gates, and BSCs that allow rapid transportation of ions across the memory cells to support the qubit interaction necessary for multi-qubit gates. Tiles are specified by their tasks, such as data storage (Data Tile), state preparation for non-Clifford gates (Ancilla Tile), error correction (EC Tile), and communication between segments (Communication Tile). At the highest layer of hierarchy, various L2 Tiles are assembled to construct two types of segments: {\em storage segments} (SS) and {\em computational segments} (CS). A segment is the largest unit that is internally connected using BSCs, and the connections between segments are carried out using optical interfaces. Each segment must contain at least one Communication Tile, one EC Tile, and several Data Tiles. The SS store qubits when they undergo \emph{no-ops}. On the other hand, CS contains qubits necessary to perform the complex non-Clifford gates, which requires the ability to prepare the magic states and support the teleportation of data. The magic state is prepared using Ancilla Tiles, which are specific to CS only. The transportation of data between segments is achieved by teleporting data through the EPR link established by the Communication Tiles of the segments. A network of optical switches~\cite{KimPTL2003} enables EPR pair generation between any pair of segments in the system. To generate an L2 logical EPR pair, an L2 CNOT consisting of 49 physical CNOT gates is first enacted using $49$ physical EPR pairs~\cite{24} and then error correction is applied to improve its fidelity~\cite{QuantRepEC}. The detailed composition of different L2 Tile types is provided in Table~\ref{label:TileComp}. An L2 Tile can contain up to 600 cells arranged in a 2-D grid, and the time to shuttle logical qubits through the Tile is defined to be $60\mu s$. 

This architecture scales by adding more qubits to the system in the form of additional segments, which demands a larger optical switch network, at nominal increase in the latency overhead of EPR pair generation. The optical switches can be connected in a tree-like hierarchy such that the height $h$ of the tree scales only logarithmically with the number of segments. We restrict the size of the optical switches to 1,000 ports~\cite{KimPTL2003}, which can connect up to 20 segments at the lowest level of the optical network tree, using 49 optical ports per L2 Communication Tile. This unique feature enables global connectivity for the entire QC, with the cross-segment communication time almost independent of distance between segments. The communication time scales with the height $h$ of switch network as $2^{h-1}$. We found that $h\leq3$ is sufficient to connect the maximum number of segments arising in our sizable benchmark circuits, and the corresponding maximum communication time is  5ms$\times2^{h-1} = 20$ms.

This globally-connected QC architecture with fast communication channels ensures rapid access to CS where computational resources for non-Clifford groups are available. This allows us to designate a finite number of CS in the overall QC to be shared across the computation. Furthermore, the physical construction of the Data and Ancilla Tiles are nearly identical (one Ancilla Tile can serve as two Data Tiles, and vice versa), so the designation between Data and Ancilla Tiles can be dynamically adjusted during the course of the computation.  The total number of segments and the allocation of Data, Ancilla, EC and Communication Tiles per SS and CS are the architectural parameters of our QC design, denoted by total number of qubits (NTQ) and segments (NSeg), number of computational segments (NCS) (NCS $\le$ NSeg), number of EC Tiles (NEC) and Communication Tiles (NComm) per segment, and the number of Data (NData) and Ancilla (NAnc) Tiles per CS and SS. Throughout our simulation analysis we assume NEC = 1 since L2 error-correction is applied sparsely to Data Tiles, and a single EC Tile can serve several L2 logical qubits. Hence, a concise description of the architecture contains (1) NCS and (2) configuration of CS specified by three numbers (NData, NAnc, NComm). For SS, we replace NAnc with 2$\times$NData. Our analysis framework will involve changing architectural parameters and studying their impact on resource-performance trade-offs.  The performance metrics consist of execution time $T_{exec}$ and failure probability $P_{fail}=1-P_{succ}$ of the circuit, where $P_{succ}$ is the probability that circuit execution yields a correct result.     

\begin{figure*}[tbp]
	\begin{center}
		\includegraphics[width=0.8\textwidth]{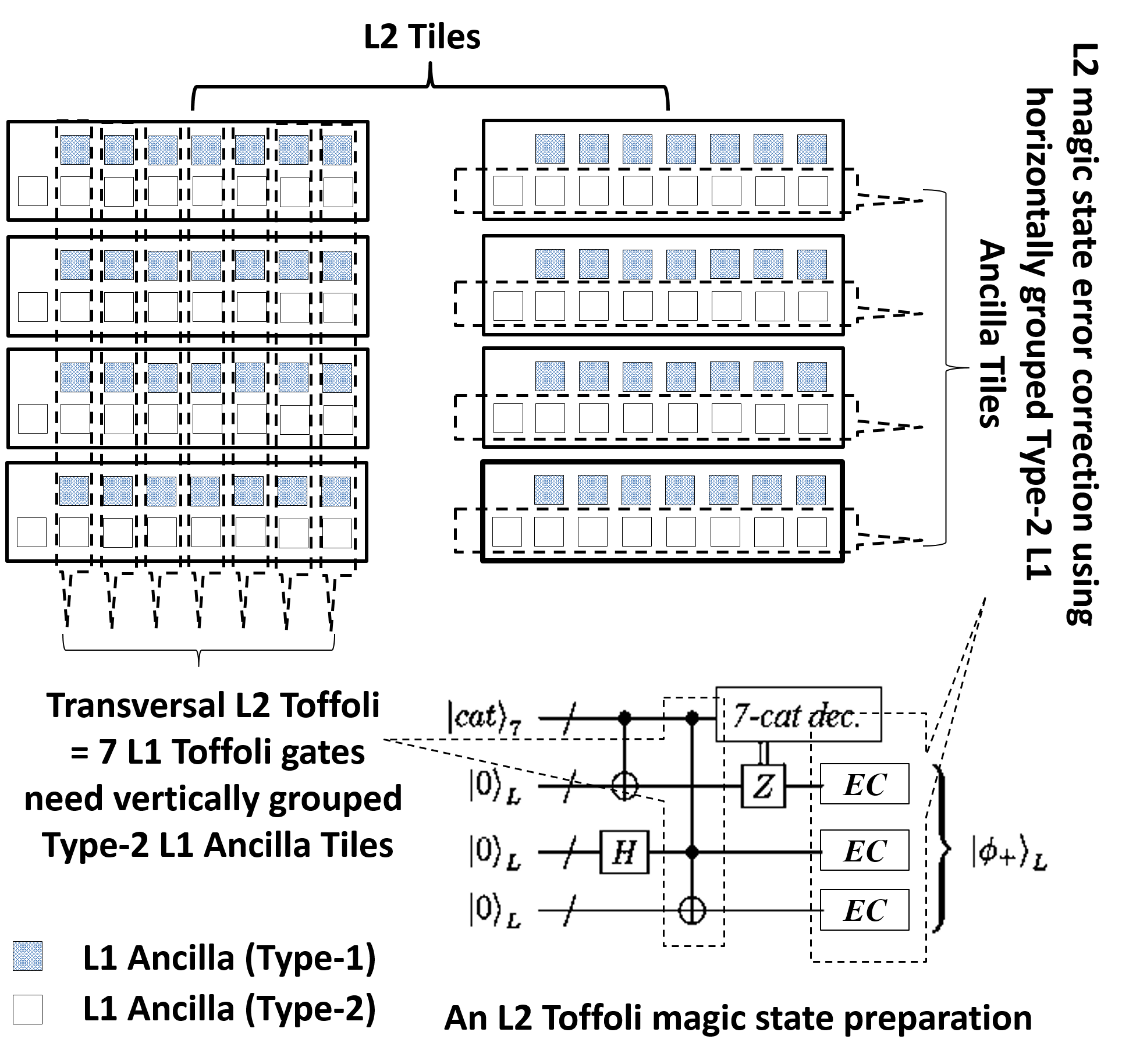}
	\end{center}
	\caption{An example demonstration of cross-layer resource optimization in L2 Toffoli magic state preparation circuit. Type-1 L1 Tiles store magic state while Type-2 L1 Tiles which initially perform transversal L1 Toffoli gates can be reallocated for L2 error correction.}
	\label{fig:CrossLayerOptimization}
\end{figure*}

\subsection{Dynamic Resource Allocation and Cross-layer Optimization}
Our system architecture provides several unique features not considered before, that provide a crucial advantage in the resource-performance optimization of the QC design. First, the L2 logical blocks are not identical instantiations of L1 logical blocks: we utilize several L1 logical blocks to construct L2 Tiles with different functionalities. 
Figure~\ref{fig:CrossLayerOptimization} shows that L2 Toffoli magic state preparation in the L2 Ancilla Tiles containing two types of L1 Ancilla Tiles. A Type-1 Ancilla will store a magic state while a Type-2 Anilla assumes multiple roles across layers of concatenation. When grouped vertically (Fig.\ref{fig:CrossLayerOptimization}(a)), Type-2 Ancilla perform transversal L1 Toffoli gates. The horizontal grouping of these Tiles (Fig.\ref{fig:CrossLayerOptimization}(b)) performs error correction for the L2 magic state. 
This ``cross-layer optimization'' allows efficient utilization of the resources for those tasks (such as error correction at L2) where common resources can be shared. Second, our system allows dynamic re-allocation of computational resources during the computation (Data vs. Ancilla Tiles) to adapt the architecture to the computational task at hand to improve the performance, analogous to reconfigurable computing using field-programmable gate arrays (FPGAs) in modern classical computing. Last, utilization of fully distributed resources (such as CS) is enabled by the global connectivity that is a unique feature of our architecture.

\section{Tool Description} \label{ToolSec}
\subsection{Design Flow and Tool Components}
\begin{figure}[tb]
	\centering
	\includegraphics[width=0.85\textwidth]{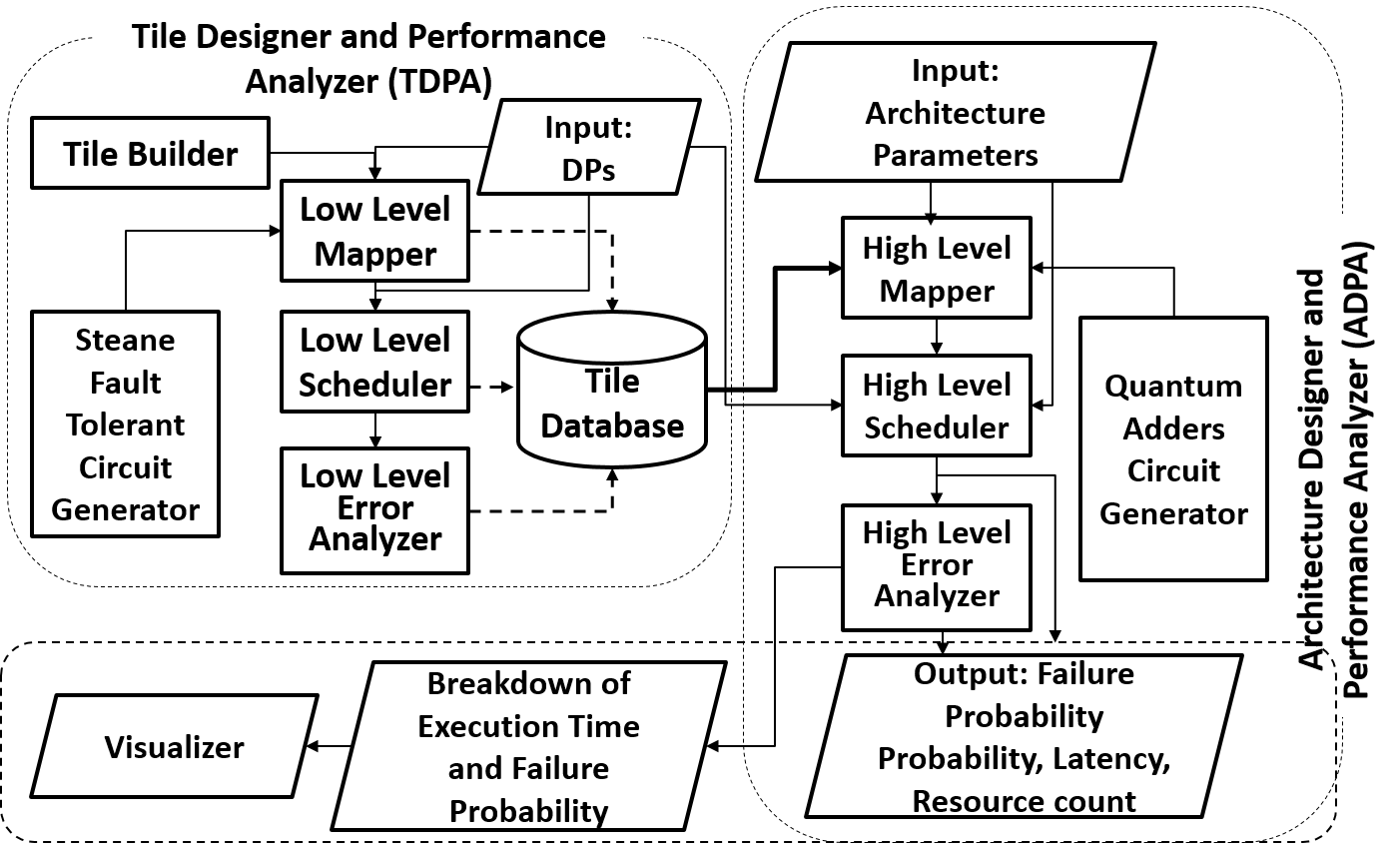}
	\caption{Main components of our toolbox}
	\label{label:Pic3}
\end{figure}

The toolbox flow shown in Fig.\ref{label:Pic3} has two main components: Tile Designer and Performance Analyzer (TDPA) and Architecture Designer and Performance Analyzer (ADPA). Both components share common critical tasks, namely \emph{mapping}, \emph{scheduling} and \emph{error analysis}, but the application of these tasks differs according to their objectives and the constraints.

\begin{center}
	\begin{table} \tbl{Composition of L2 Tile\label{label:TileComp}}{
			\begin{tabular}[c]{ |c|c|c| }	
				\hline
				Tile Type & L1 Tiles & Physical qubits\\ \hline \hline 
				Data  & $7$ & $154$    \\ \hline
				Ancilla  & $15$ & $330$   \\ \hline
				Error Correction & $15$ & $330$  \\ \hline
				Communication  & $22(=7 + 15)$ & $484 + 49$   \\ \hline
			\end{tabular}}
	\end{table}
\end{center}

\subsubsection{TDPA}
TDPA works in the back end of the tool and simulates the fault-tolerant construction of the logical qubit operations using specified DPs. It builds Tiles using \textbf{Tile Builder} by allocating sufficient qubits that can perform the operations specified in \textbf{Steane Fault-Tolerant Circuit Generator}, and maps qubits in the circuit to the physical qubits in the Tile using \textbf{Low Level Mapper}. Then, \textbf{Low Level Scheduler}  generates the sequence of quantum gate operations to be executed in the circuit, including transversal gates, magic state preparation for non-Clifford gates, error correction and EPR pair generation. Each logical operation is broken down into constituent physical operations, whose performance is simulated on the Tile by adding up the execution time of each gate subject to circuit dependencies and resource constraints. \textbf{Low Level Error Analyzer} computes the failure probability of the specified fault-tolerant quantum gates based on the scheduled circuit, by counting the number of ways in which physical errors can propagate to cause a logical error in the qubit~\cite{AGP,ahsan2013performance}. The Tile parametrized by DP and the computed performance metrics is stored in \textbf{Tile Database}. Table~\ref{label:T1} shows the performance of the unified L2 Tile (which can act as Data, EC, Ancilla and Communication Tile) computed by the TDPA for baseline DPs given in Table~\ref{label:T0}.

\subsubsection{ADPA}
ADPA is the front end of the tool that interfaces with the user. It takes architecture parameters specified by the user (e.g., NCS, NEC and NComm) as inputs and (1) builds and connects segments using Tiles supplied by TDPA to implement benchmark application on hardware configuration, and (2) evaluates performance of the benchmark for given architecture parameters. First, \textbf{Quantum Circuit Generator} generates the benchmark circuits from the given algorithms (QCLA, QRCA and AQFT). Then, \textbf{High Level Mapper} maps logical qubits to Tiles in the segments, maximizing the locality by analyzing their connectivity patterns in the circuit, assigning frequently interacting qubits to the Tiles in the same segment. This is achieved by solving an \emph{optimal linear arrangement problem} using an efficient graph-theoretic algorithm~\cite{11} to generate the initial map of the Data Tiles in the segments. Using this map, \textbf{High Level Scheduler} generates the sequence of gates for the circuit execution by solving the standard \emph{resource-constraint scheduling problem} in which resources and constraints are given by architecture parameters. Scheduler minimizes the execution time by reducing the \emph{circuit critical path} through maximum utilization of available resources (Ancilla and Communication Tiles) in the segments. The non-Clifford gates require operands to be available in the same CS before being scheduled. Therefore the operand located in remote Data Tiles needs to be teleported into the local Data Tile of the CS, while Ancilla Tiles prepare the magic state for execution. 
NCS determines how many non-Clifford gates can be scheduled in parallel, while NComm determines how quickly Tiles can be teleported across the segments. Therefore the delays in gate scheduling depend mainly on architecture parameters. The critical path of the circuit consists of these delays and the gate execution time. The complete list of latencies arising due to insufficient resources and architecture configuration is: 
\begin{itemize}
	\item \emph{Ancilla Delay ($D_{ANC}$)}: Delay due to the magic state preparation (fewer NCS or fewer Ancilla Tiles per segment)
	\item \emph{Shuttling Delay ($D_{SHUT}$)}: Delay due to the transportation of operand qubits of the gate, through BSC inside the segment
	\item \emph{Tel Delay ($D_{TEL}$)}: Delay due to the logical EPR pair generation for communication (Fewer NComm)
	\item \emph{Cross-Seg-Swap Delay ($D_{SWP}$)}: Delay due to the cross-segment swapping (fewer NComm or large number of smaller segments)
\end{itemize}

The Scheduler also minimizes $P_{fail}$ by scheduling error correction on Data Tiles at regular intervals when they sit idle (\emph{no-op}). Once a complete schedule of logical operations is obtained, \textbf{High Level Error Analyzer} computes the overall $P_{fail}$. Since we cannot correct for logical failure of the operation, Error Analyzer simply computes $\prod_{i=1}^{i=N}(1-P_{Li})$, where $P_{Li}$ is the failure probability of the $i$-th logical gate and $N$ is the total number of logical operations in the circuit. Error Analyzer also tracks the operational source of each $P_{Li}$ so that $1-P_{succ}$ can be broken down into the following important noise components:

\begin{itemize}
	\item \emph{Shuttling Error} $P_{SHUT}$: Errors due to the qubit shuttling through noisy BSC
	\item \emph{Teleportation Noise} $P_{TEL}$: Errors due to the infidelity of an EPR pair for communication 
	\item \emph{Memory Noise} $P_{MEM}$: Errors due to the fidelity degradation of qubit during \emph{no-op}
	\item \emph{Gate Noise} $P_{GATE}$: Errors due to the noisy quantum gates and measurements
\end{itemize}
\begin{center}
	\begin{table} \tbl{L2 Tile Performance Numbers. The L1 Error Correction takes 687$\mu s$ and fails with probability $1.66 \times 10^{-10}$. \label{label:T1}}{

			\begin{tabular}[c]{ |c|c|c| }
				
				\hline
				Logical Operation & \textbf{$T_{exec} (\mu s)$} & \textbf{$1-P_{succ}$}  \\ \hline \hline 
				Pauli (X, Z)  &1 & $1.15\times 10^{-18}$    \\ \hline
				Hadamard  &4 & $1.15\times 10^{-18}$   \\ \hline
				CNOT &$10$ & $4.74\times 10^{-18}$    \\ \hline
				Transversal (Bitwise) Toffoli  &$4,210$ & $1.1\times 10^{-17}$   \\ \hline
				7-qubit Cat-State prep.  & $6,500$ & $3.75\times 10^{-18}$    \\ \hline
				Measurement  & $11,900$ & $6.14\times 10^{-17}$    \\ \hline 
				L2 Error Correction  & $48,900$ & $4.58\times 10^{-16}$   \\ \hline
				State prep. ($|\overline{0}\rangle$,$|\overline{+}\rangle$)  & $34,500$ & $1.6\times 10^{-16}$   \\ \hline
				State prep. ($T|\overline{+}$)  & $78,100$ & $4.23\times 10^{-16}$   \\ \hline
				EPR pair generation  & $T_{gen}+50,800$ & $1.08\times 10^{-11}$   \\ \hline
			\end{tabular}}
	\end{table}
\end{center}

\subsection{Splitting Performance Metrics}

\begin{figure}[tb]
	\centering
	\includegraphics[width=0.8\textwidth]{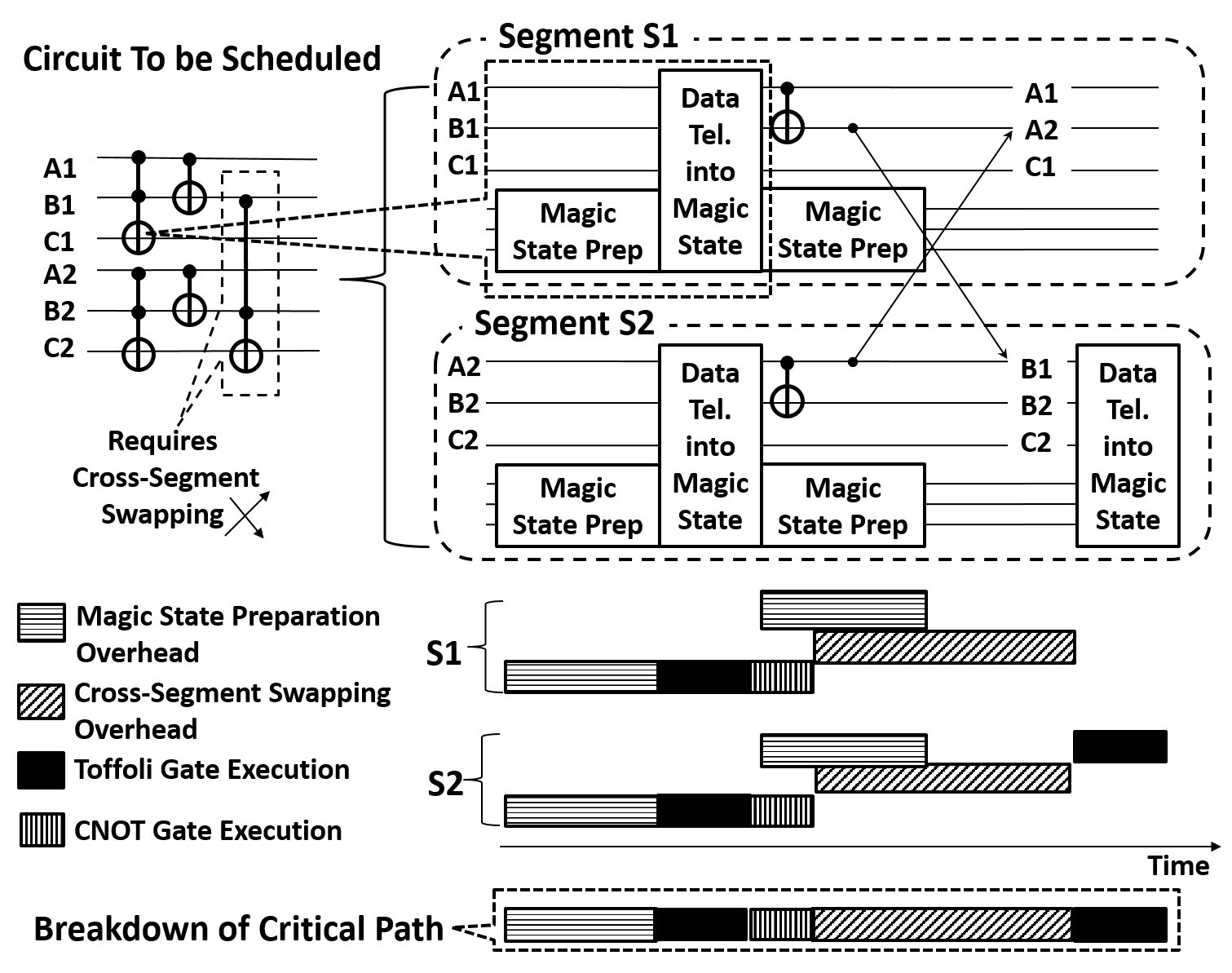}
	\caption{An example shows tracking of latency overheads comprising critical path of the circuit.}
	\label{label:Pic2}
\end{figure}

Our tool can output the constituents of $T_{exec}$ by keeping track of the different types of latency overhead comprising the critical path of the quantum circuit execution. Our iterative scheduler selects a gate for execution during each iteration and updates the critical path. When a gate is selected for execution, the operand qubits should be available for computation otherwise it will be delayed. If we let $T_{start}$ be the time at which the gate was executable but operands were available at $T'_{start}$ where $T'_{start} \ge T_{start}$, then the time at which the gate execution is complete is given by  $T_{finish} = T'_{start} + T_{exec}$. The gate execution selected in scheduling iteration $i$ will update the critical path if $T_{finish} > T_{TotalExec}^{i-1}$ where $T_{TotalExec}^{i-1}$ is the total execution time (the length of critical path) computed in iteration $i-1$.   

When the gate lies on the critical path, the Scheduler computes $\Delta T = T_{finish} - T_{TotalExec}^{i-1}$ and $\Delta D = T'_{start} - T_{start}$ which can be broken down as $\Delta D = D_{ANC} + D_{SHUT} + D_{TEL} + D_{SWP}$. To define the components of the critical path in iteration $i-1$, we split $T_{TotalExec}^{i-1}$ into its components as: $T_{TotalExec}^{i-1} = T_{ANC}^{i-1}+T_{SHUT}^{i-1}+T_{TEL}^{i-1}+T_{SWP}^{i-1}+T_{GATE}^{i-1}$. For iteration $i$, these components are updated as follows: 

\begin{itemize}
	\item \emph{Ancilla Preparation Overhead}:
	
	$ T_{ANC}^i = T_{ANC}^{i-1} + \frac {D_{ANC}}{\Delta D + T_{exec}} \times \Delta T$    
	
	\item \emph{L2 Shuttling overhead}: 
	
	$ T_{SHUT}^i = T_{SHUT}^{i-1} + \frac {D_{SHUT}}{\Delta D + T_{exec}} \times \Delta T$ 
	
	\item \emph{Teleportation Overhead}:
	
	$ T_{TEL}^i = T_{TEL}^{i-1} + \frac {D_{TEL}}{\Delta D + T_{exec}} \times \Delta T$   
	
	\item \emph{Segment Swap Overhead}:
	
	$ T_{SWP}^i = T_{SWP}^{i-1} + \frac {D_{SWP}}{\Delta D + T_{exec}} \times \Delta T$

	\item \emph{Gate overhead}: 
	
	$ T_{GATE}^i = T_{GATE}^{i-1} + \frac {T_{exec}}{\Delta D + T_{exec}} \times \Delta T$ 
	
\end{itemize}

\noindent In the case of the critical path, we also update the total execution time $T_{TotalExec}^{i} = T_{finish}$. An example breakdown of execution time is shown in Fig.~\ref{label:Pic2}, where magic state preparation for Toffoli gates and cross-segment swapping delay the execution of gates in turn and comprise the bulk of the critical path. Our tool can also decompose the failure probability $P_{fail}$  into its components since Error Analyzer tracks noise sources from the schedule. For any operation type \emph{op}, we compute  $P_{fail}^{op} =1- \prod_{i=1}^{i=n^{op}}(1-P_{Li}^{op})$, where ${op}$ can be \emph{shuttling}, \emph{memory}, \emph{teleportation} or \emph{gate}, and $n^{op}$ and $P_{Li}^{op}$ are the total operation count and failure probability for \emph{op}, respectively.  

\begin{figure*}[tb]
	\centering
	\includegraphics[width=0.7\textwidth]{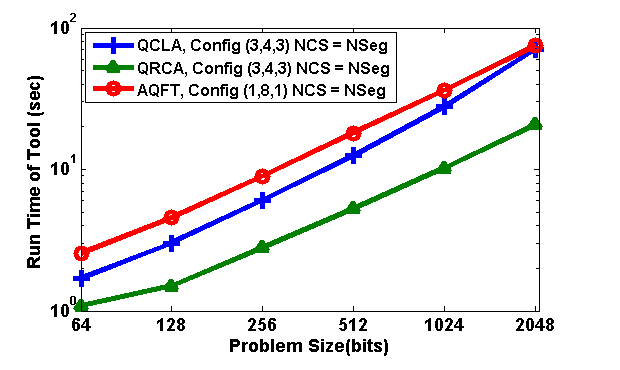}
	\caption{The running time of ADPA (excluding Visualizer) as a function of benchmark application size. With full Visualization, the running times increases no more than seven times.}
	\label{label:ToolRunTime}
\end{figure*}

\subsection{Tool Validation and Performance}
Individual components of the tool can easily be verified for correctness by running these for known circuits and comparing their output with anticipated results. Overall validation can be performed by using visualization and the breakdown of performance metrics for different types of benchmarks. Tool efficiency mainly arises from taking advantage of the repetitive nature of fault-tolerant procedures and circuit breakdown of universal quantum gates. Performance of low-level circuit blocks is pre-computed and stored in the database, and used for simulating the behavior of high-level circuits. For instance TDPA can be run offline to generate parametrized Tiles, which are used to efficiently run components of ADPA such as High Level Scheduler, Error Analyzer and Visualizer. Similarly, the initial mapping of L2 qubits on these Tiles is generated from a computationally intensive optimization algorithm ~\cite{11}, but once generated, can be efficiently processed by the High Level Scheduler to generate subsequent gate schedules. Thus, we can run High Level Mapper offline as well. Consequently, the running time of the tool is decided by that of High level Scheduler and Error Analyzer, which mainly depends on architecture resources and benchmark size. The results discussed in Section~\ref{SimSec} show that the performance improvement saturates once resource investment exceeds a certain value, and the maximum size of the overall system we have to simulate is mainly dictated by the size of the application circuit. 

Figure~\ref{label:ToolRunTime} shows the running time of the tool as function of circuit size. The data is collected by running the tool on a computer system containing Intel$^{\mathrm{(R)}}$Core$^{\mathrm{(TM)}}$ $i3$ 2.4GHz processor and 2GB RAM. To incorporate the dependency of tool running time on the available architecture resources, we choose the configuration containing maximum resources in order to obtain typical worst-case running time of the tool. In this configuration, we allocate maximum Ancilla and Communication Tiles per Data Tile and allow all Segments to act as Computational Segments. Under these conditions, Fig.\ref{label:ToolRunTime} shows that the performance-simulation of 2,048-bit circuit can be completed in less than 1.5 minutes. Thanks to the efficiency in the performance simulation, our tool can explore a large QC design space in a reasonable amount of time.              

\section{Simulation Results} \label{SimSec}

\begin{figure*}[tb]
	\centering
	\includegraphics[width=1\textwidth]{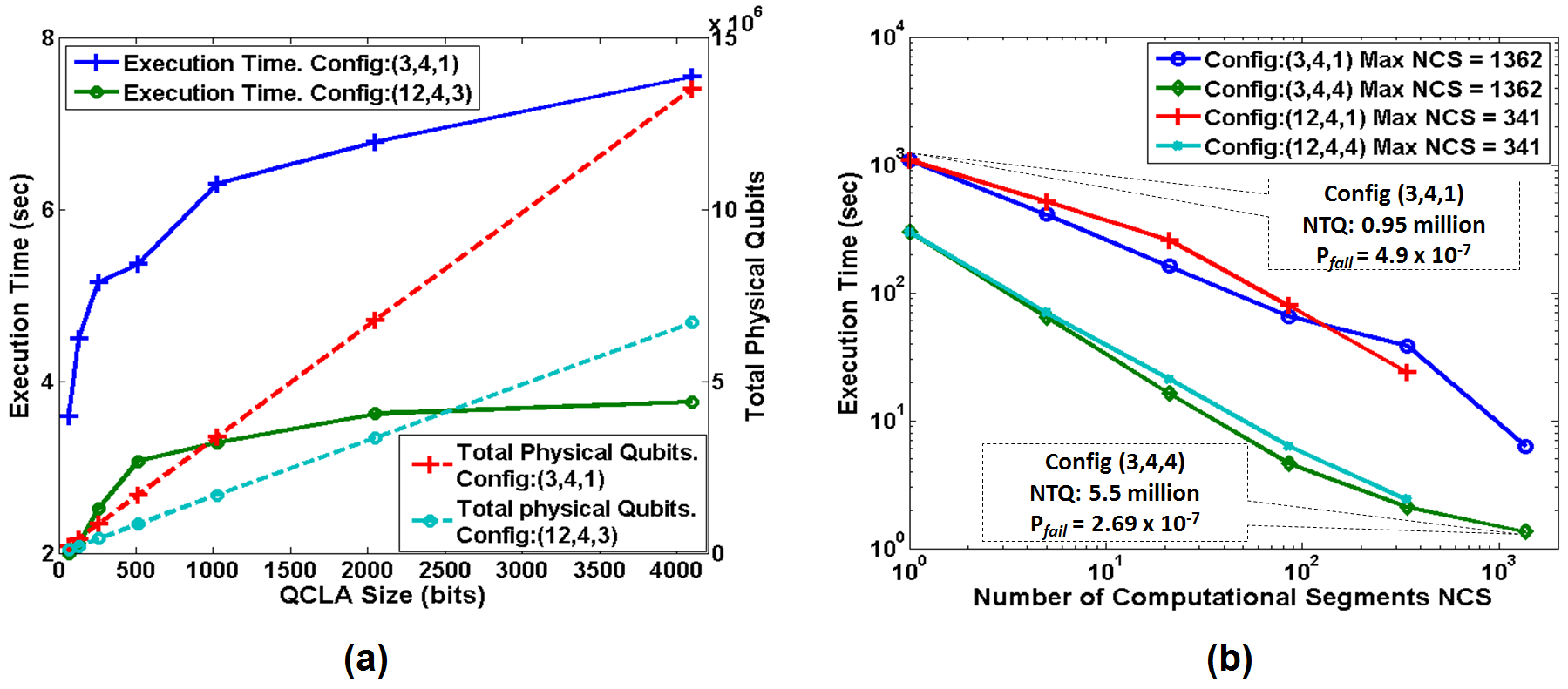}
	\caption[Resource-performance analysis of QCLA]{QCLA execution time (a) scaling under no constraints on physical resources, $T_{exec}$ and total physical qubits (NTQ) consumed are plotted as a function of benchmark size, NCS = NSeg.(b) variation with NCS for different NComm, showing trade-offs between resources and $T_{exec}$}
	\label{label:QCLA_all}
\end{figure*}

\begin{figure}[tb]
	\centering
	\includegraphics[width=1\textwidth]{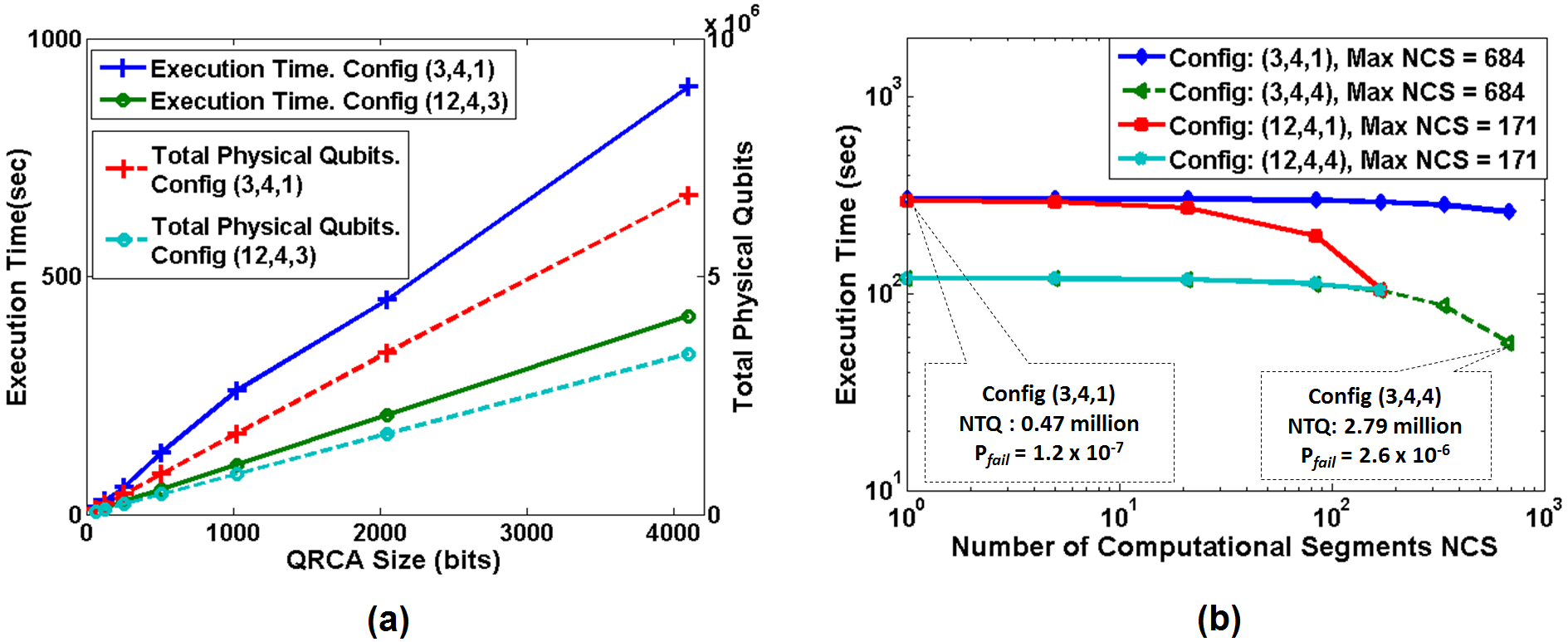}
	\caption[Resource-performance analysis of QRCA]{QRCA execution time (a) scaling under no constraints on physical resources, $T_{exec}$ and total physical qubits (NTQ) consumed are plotted as a function of benchmark size, NCS = NSeg.(b) variation with NCS for different NComm, showing trade-offs between resources and $T_{exec}$}
	\label{label:QRCA_all}
\end{figure}

\begin{figure}[tb]
	\centering
	\includegraphics[width=1\textwidth]{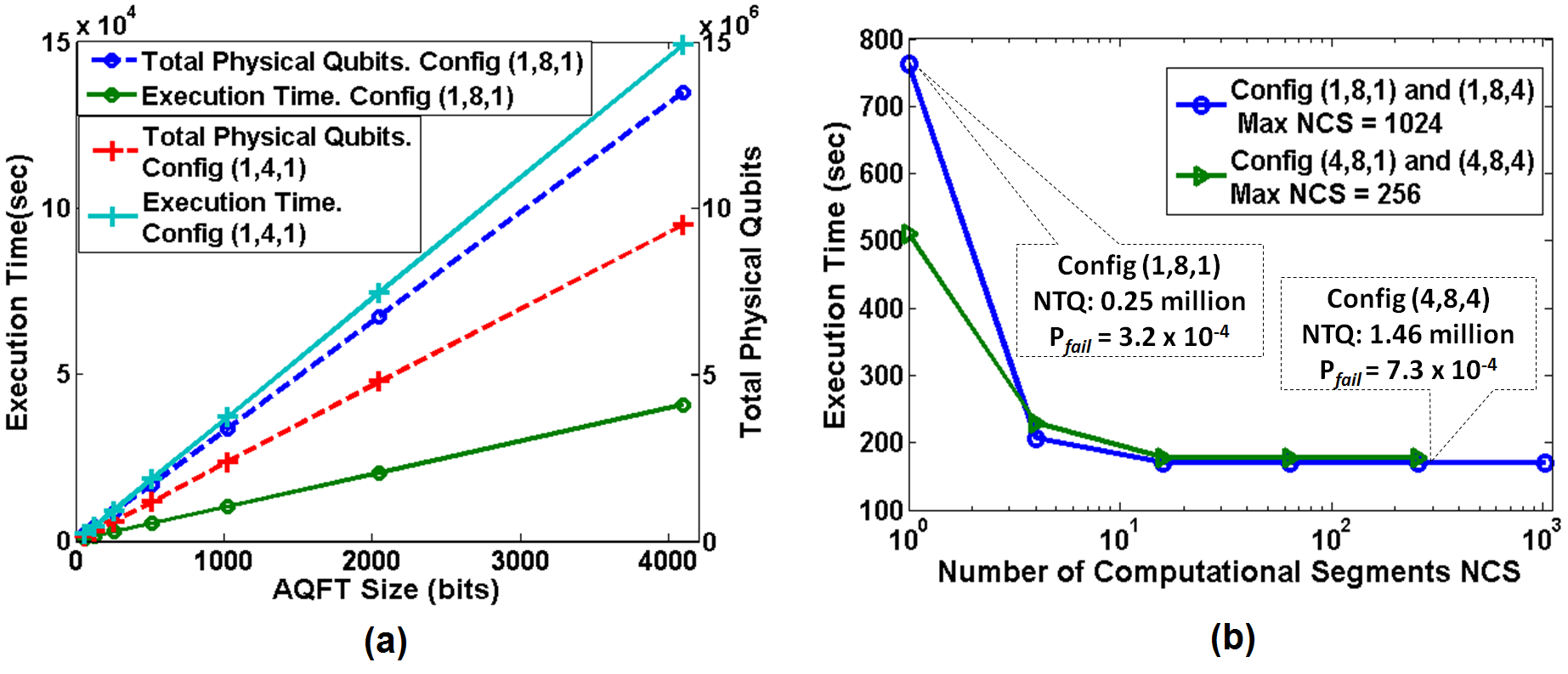}
	\caption[Resource-performance analysis of AQFT]{AQFT execution time (a) scaling under no constraints on physical resources, $T_{exec}$ and total physical qubits (NTQ) consumed are plotted as a function of benchmark size, NCS = NSeg.(b) variation with NCS for different NComm, showing trade-offs between resources and $T_{exec}$}
	\label{label:QFT_all}
\end{figure}
	
We first analyze the relationship between resources (qubits) and performance as a function of benchmark size for scalability. We consider the system architecture \emph{resource-performance} (RP) \emph{scalable} if the increase in resources necessary to achieve the expected behavior of the performance (execution time) grows linearly with the size of the benchmark, while maintaining $P_{succ} \sim O(1)$. In the absence of hardware resource constraints, the expected execution time for the QRCA and AQFT grows linearly, while that for QCLA grows logarithmically, as the problem size grows. The execution time could grow much more quickly in the presence of resource constraints, in which case the system is not considered RP scalable.

In the first step we present a set of simulations to quantify the RP scalability of the proposed MUSIQC architecture for benchmark circuits and analyze the constituents of performance metrics. In the next step, we study the impact of limited resources and architecture parameters on the performance of fixed size benchmarks. This will provide guidelines to find an optimized design under limited resources. In the last set, optimum designs are obtained under resource constraints, for effectively executing the benchmark circuits.   

\subsection{Resource-Performance Scalability}

\begin{table}
	\tbl{Failure Probability $P_{fail}$ for corresponding data points of in Fig.~\ref{label:QCLA_all}(a),\ref{label:QRCA_all}(a),\ref{label:QFT_all}(a)
	\label{label:Tab_Failure_Scale}}{
	\centering
	\includegraphics[width=0.7\textwidth]{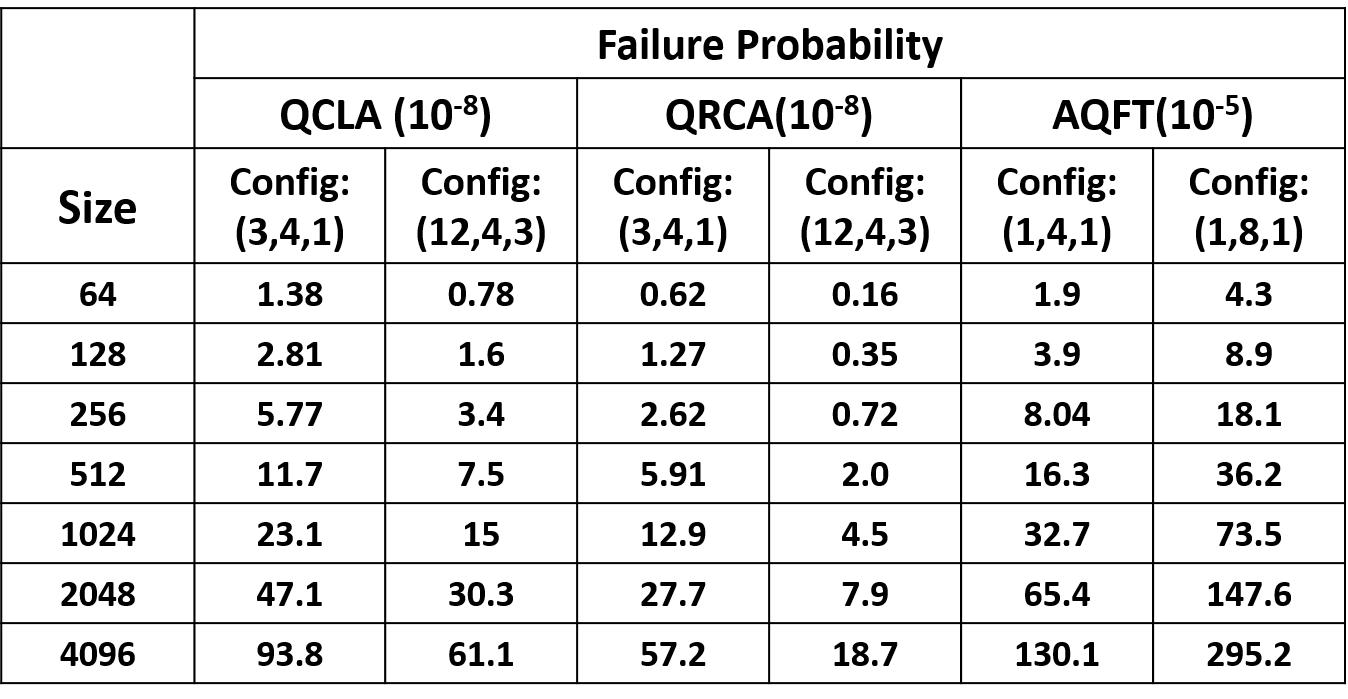}}
\end{table}

\begin{table}
	\tbl{The breakdown of Performance metrics: $T{exec}$ of Fig.~\ref{label:QCLA_all}(a),\ref{label:QRCA_all}(a),\ref{label:QFT_all}(a) and $P_{fail}$ of Table~\ref{label:Tab_Failure_Scale} for 1,024-bit benchmark. NCS = NSeg \label{label:BreakDown}}{
	\centering
	\includegraphics[width=.8\textwidth]{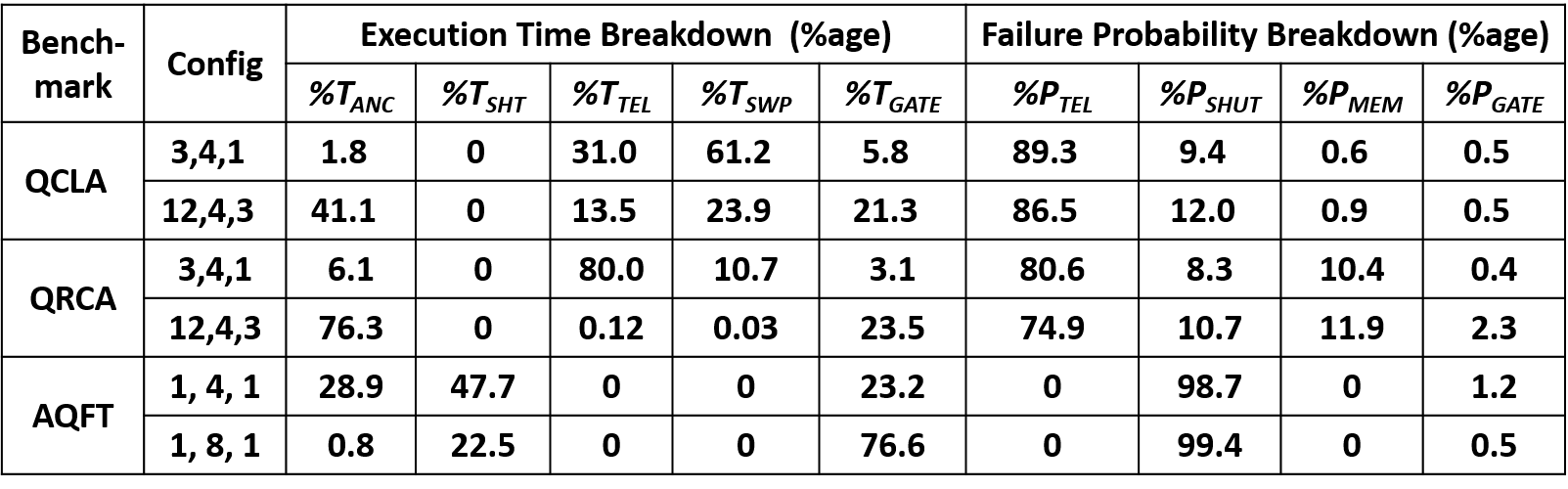}}
\end{table}

Fig.~\ref{label:QCLA_all}(a),~\ref{label:QRCA_all}(a) and ~\ref{label:QFT_all}(a) show $T_{exec}$ and the total number of physical qubits (NTQ) plotted against benchmark size for QCLA, QRCA and AQFT respectively, and corresponding $P_{fail}$ values are shown in Table~\ref{label:Tab_Failure_Scale}. We consider two architecture configurations (NData, NAnc, NComm)=(3,4,1) and (12,4,3) for the adders, and configurations (1,4,1) and (1,8,1) for AQFT. When benchmark size increases by factor of $x$, we expect $P_{fail}$ to increase by at least the same amount (total logical gate operations increase $x$-fold) while execution time scales as the depth of the circuit. Indeed Table~\ref{label:Tab_Failure_Scale} confirms this trend in $P_{fail}$ for all benchmarks. On the other hand, as problem size doubles, both $T_{exec}$ for QRCA and AQFT increase 2 fold [linear curve for $T_{exec}$ in Fig.~\ref{label:QRCA_all}(a) and ~\ref{label:QFT_all}(a)]. For QCLA,  $T_{exec}$ increases roughly by a constant amount [logarithmic curve in Fig.~\ref{label:QCLA_all}(a)] as expected. Since this performance is achieved for the same increase in total number of qubits as that of the problem size, our architecture shows RP scalability.  

By demonstrating RP scalability for two different architecture configurations with NCS = NSeg, we have presented varying performance levels. The impact of these configurations can be understood by analyzing the breakdown of performance metrics in Table~\ref{label:BreakDown}. The significant contribution of the overhead $T_{ANC}$ for adder configuration (12,4,3) and AQFT configuration (1,4,1) shows that magic state preparation is the dominant component of $T_{exec}$ due to insufficient Ancilla Tiles in CS. This overhead can be substantially reduced either by increasing NAnc (configuration (1,8,1) for AQFT) or by increasing the ratio of NData to NAnc (configuration (3,4,1) for adders).  However, the configuration (3,4,1) exposes EPR pair generation overhead captured by $T_{TEL}$ and $T_{SWP}$. This is due to the large number of cross-segment CNOT gates and qubit swapping operations required to bring all Toffoli operands to the same segment. Hence for adders, frequent cross-segment communication explains the higher contribution of teleportation error ($P_{TEL}$) in the failure probability. For AQFT, configuration (1,8,1) highlights shuttling overhead $T_{SHUT}$ as $T$ gates comprise bulk of the operations. The scheduling of $T$ gates in the long approximation sequence leads to a large number of interactions between Data and Ancilla Tiles through BSC in the segment. This intensive localized communication makes ($P_{SHUT}$) the only noticeable component of $P_{fail}$. In conclusion, we have shown RP scalability of the architecture when performance is bottlenecked by different hardware constraints for various benchmarks. This shows that the architecture can utilize additional resources efficiently to achieve adequate performance when running quantum circuits of larger sizes.    

\subsection{Resource-Performance Trade-offs}

\begin{figure}[tb]
	\centering
	\includegraphics[width=1\textwidth]{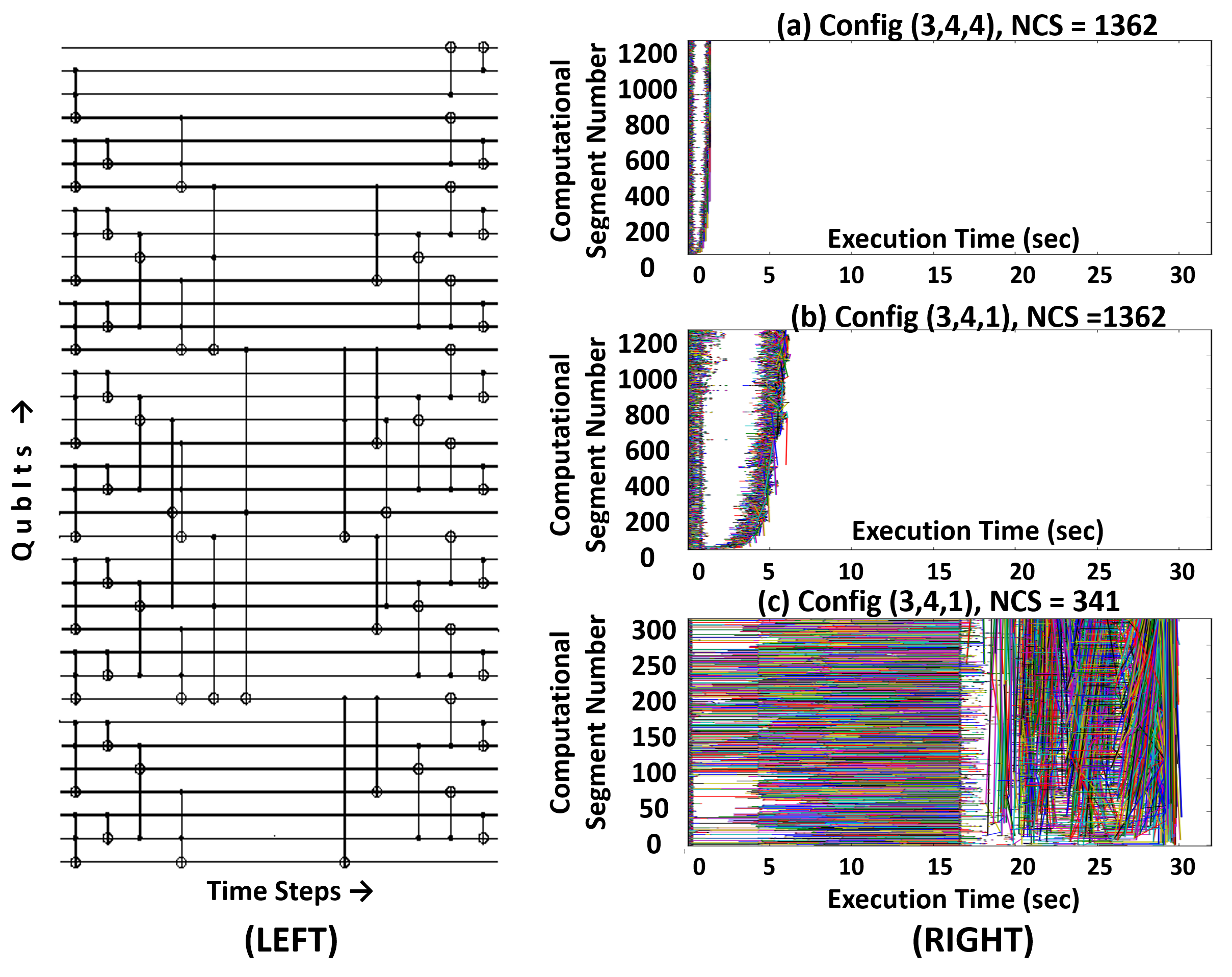}
	\caption{(LEFT) Sample QCLA circuit, and (RIGHT) visual representation of latency overhead for Computational Segments of 1024-bit QCLA architecture with (a) configuration (3,4,4), NCS = NSeg = 1362, (b) configuration (3,4,1), NCS = NSeg = 1362 and (c) configuration (3,4,1), NCS = 341.}
	\label{label:Visual1}
\end{figure}

Now that we have quantified RP scalability, we examine the impact of reduced resources on the performance by constraining architecture parameters. We fix the size of the benchmarks to 1,024 bits and vary NComm, NCS and the configuration to observe changes in $T_{exec}$. Fig.~\ref{label:QFT_all}(b) shows that for AQFT, (1) $T_{exec}$ does not change with NComm since it is not restricted by cross-segment communication resources, and (2) $T_{exec}$ initially decreases sharply as  NCS increases, but flattens when NCS reaches $\sim$ 16, mainly due to insufficient parallelism in the circuit. The QRCA curves in Fig.\ref{label:QRCA_all}(b) remain unchanged as NCS increases until NCS approaches NSeg where $T_{exec}$ shows noticeable decline. However, the overall decrease remains within a factor of only about 3, due to serial nature of the circuit dependencies. By comparing curves for different configurations [(12,4,1) vs. (3,4,1) for QRCA and (1,8,x) vs (4,8,x) for AQFT], we find that clustering more Data Tiles in the CS generally reduces $T_{exec}$ due to fewer delays in cross-segment operand swapping. 

Fig.~\ref{label:QCLA_all}(b) shows that $T_{exec}$ of QCLA decreases exponentially with NCS. In contrast to QRCA, the large number of concurrently executable Toffoli gates in QCLA demands much higher NCS. Furthermore, $T_{exec}$ also decreases with higher NComm as the large number of cross-segment teleportations consume more communication resources. Fig.~\ref{label:Visual1}(a)-(c) provide pictorial description of these trends generated through our visualization tool. The visualization highlights different types of latencies arising from the resource constraints in scheduling within each CS. The visualization tool draws a line, in execution time (horizontal axis) -- qubit location (vertical axis) plane, between each point where a logical qubit in the circuit requires additional resources  to proceed to the next step (such as a magic state for non-Clifford group gates prepared in an Ancilla Tile, or EPR pairs for teleportation in Communication Tiles), and a point where the required resource becomes available. As a consequence, the horizontal lines represent delay in magic state preparation, while non-horizontal lines indicate delays in cross-segment teleportation. The corresponding latencies can be derived by projecting these lines on the horizontal axis. When sufficient NCS and NComm resources are provided as in Fig.~\ref{label:Visual1}(a), there is little delay and the circuit execution is fast. However, when NComm is reduced from 4 to 1 as in Figure~\ref{label:Visual1}(b) teleportation latency caused a 4.5-fold increase in $T_{exec}$. The same amount of increase occurs in Fig.\ref{label:Visual1}(c) when NCS is reduced from 1,362 to 341. Long horizontal lines indicate delays due to fewer Ancilla Tiles available for Toffoli magic state preparation, which increases the  overall $T_{exec}$ by another factor of about 6.

By comparing Figure~\ref{label:QCLA_all},\ref{label:QRCA_all},\ref{label:QFT_all}, it is easy to conclude that QCLA is the most resource hungry benchmark while AQFT is the least. We also note that $P_{fail}$ values do not tend to improve substantially when we provide more architecture resources. It can be shown that a substantial decrease in $P_{fail}$ can be achieved by improving the relevant device parameters (DPs) that contributes to the dominant noise sources~\cite{Ahsan2015}. These sources can correctly be identified once we have invested sufficient resources for scheduling error-correction and chosen optimized architecture configuration that minimizes $T_{exec}$. In the following subsection, we concentrate on $T_{exec}$ only and return to optimizing $P_{fail}$ in the last subsection. 

\subsection{Performance Scaling under Limited Resources}
The increase in $T_{exec}$ due to constrained resources can be compensated to some extent by designing an optimized architecture that allocates more resources towards the root cause that limits the performance. The choice of optimized architecture configuration varies across benchmarks, since both performance bottlenecks and resource utilization depend strongly on the structure of the application circuit. By plotting optimized $T_{exec}$ against benchmark size using a fixed resource budget, we can determine (1) the largest circuit size which can be scheduled and executed, (2) the trend for the optimized performance as a function of problem size, and (3) the choice of configuration which generally obtains the optimized performance for the specified benchmark circuit. To adequately obtain these insights, we consider an example where we restrict our total qubit resource (NTQ) to 1.5 million physical qubits. We also restrict the number of physical qubits contained in the segment; small segment (SSeg) will contain up to 5,000 physical qubits, while large segment (LSeg) will contain up to 10,000 physical qubits. For each benchmark we obtain two plots, one for each segment size. 

\begin{figure}[tb]
	\centering
	\includegraphics[width=1\textwidth]{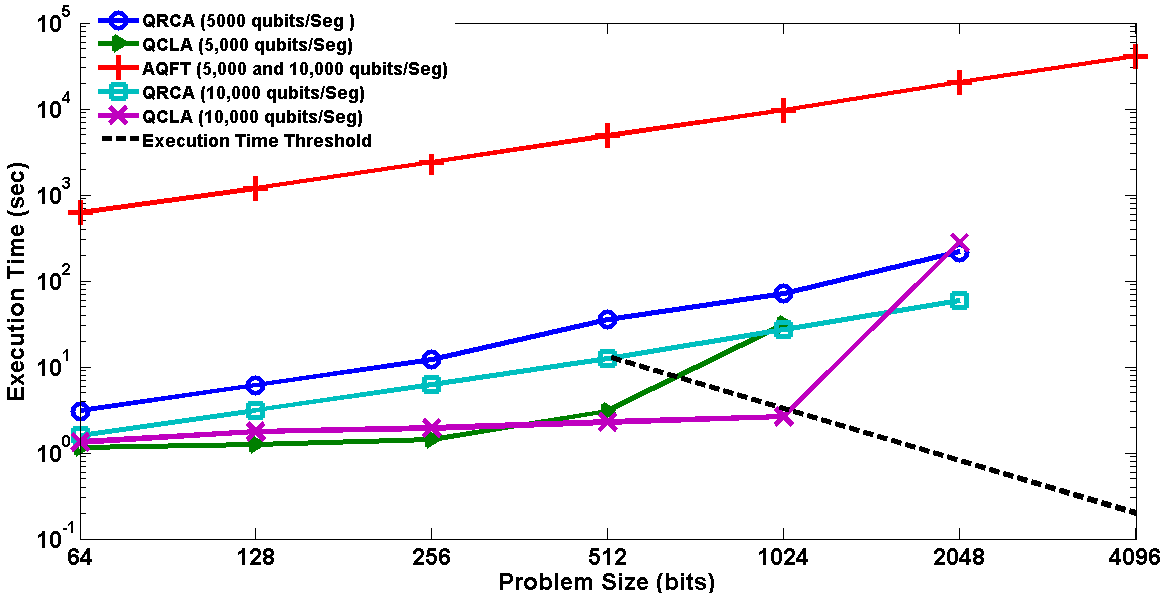}
	\caption{$T_{exec}$ for optimized architectures plotted against benchmark size for different segment sizes. The resource budget is 1.5 million physical qubits. Optimized architecture configurations are shown in Table~\ref{label:Optimum_Arch}.}
	\label{label:Pic_Problem_Scale}		
\end{figure}

Figure~\ref{label:Pic_Problem_Scale} shows $T_{exec}$ of running the benchmark circuits on this system, as a function of problem size. The $T_{exec}$ for each data point was minimized by tool-assisted search through all feasible combinations of architecture configurations and NCS parameters. These optimized architecture designs are shown in Table~\ref{label:Optimum_Arch}. It is interesting to compare Ancilla and Communication Tiles invested in the optimized designs. For example, the configurations (19,12,6) and (8,8,2) for QRCA contains a higher Ancilla-to-Communication Tiles ratio, as compared to (30,8,5) and (5,4,5) for QCLA. This indicates that in contrast to the QRCA case, both Ancilla and Communication Tiles are equally vital for the QCLA performance. The AQFT architecture (1,8,1) with sufficient NCS is a natural choice, since the only relevant resource for this circuit is the large number of Ancilla Tiles to schedule the long sequence of $T$ gates necessary to approximate the small-angle rotations.

We observe that regardless of the segment size, $T_{exec}$ of the least resource demanding benchmark of the three, namely AQFT, scales perfectly with problem size at least up to 4,096 bits. This is due to the fact that the optimized configuration for AQFT can easily be met within the qubit resource budget. However, in the case of adders, segment size is an important parameter that impacts the performance. Larger segments open up a greater search space for optimizing design selection for resource-intensive circuits. For both QCLA and QRCA, $T_{exec}$ scales better for larger segments. For smaller segments, QCLA performance shows significant degradation as the problem size begins to increase. The logarithmic depth of the most resource-intensive benchmark (QCLA) is restricted to 256-bit and 1024-bit for SSeg and LSeg, respectively. After that, $T_{exec}$ shows a sudden rise with problem size and even surpasses the corresponding QRCA performance, as lack of resources leads to substantial delays in executing the parallel gate operations that enable the logarithmic-depth adder. The largest adder benchmark that can be scheduled on this hardware is the 2,048-bit adder, where the performance is substantially slower and QRCA outperforms QCLA by factor of 4.

\begin{table}
	\tbl{Optimized Architecture Configurations for Fig.\ref{label:Pic_Problem_Scale}
	\label{label:Optimum_Arch}}{
	\centering
	\includegraphics[width=0.75\textwidth]{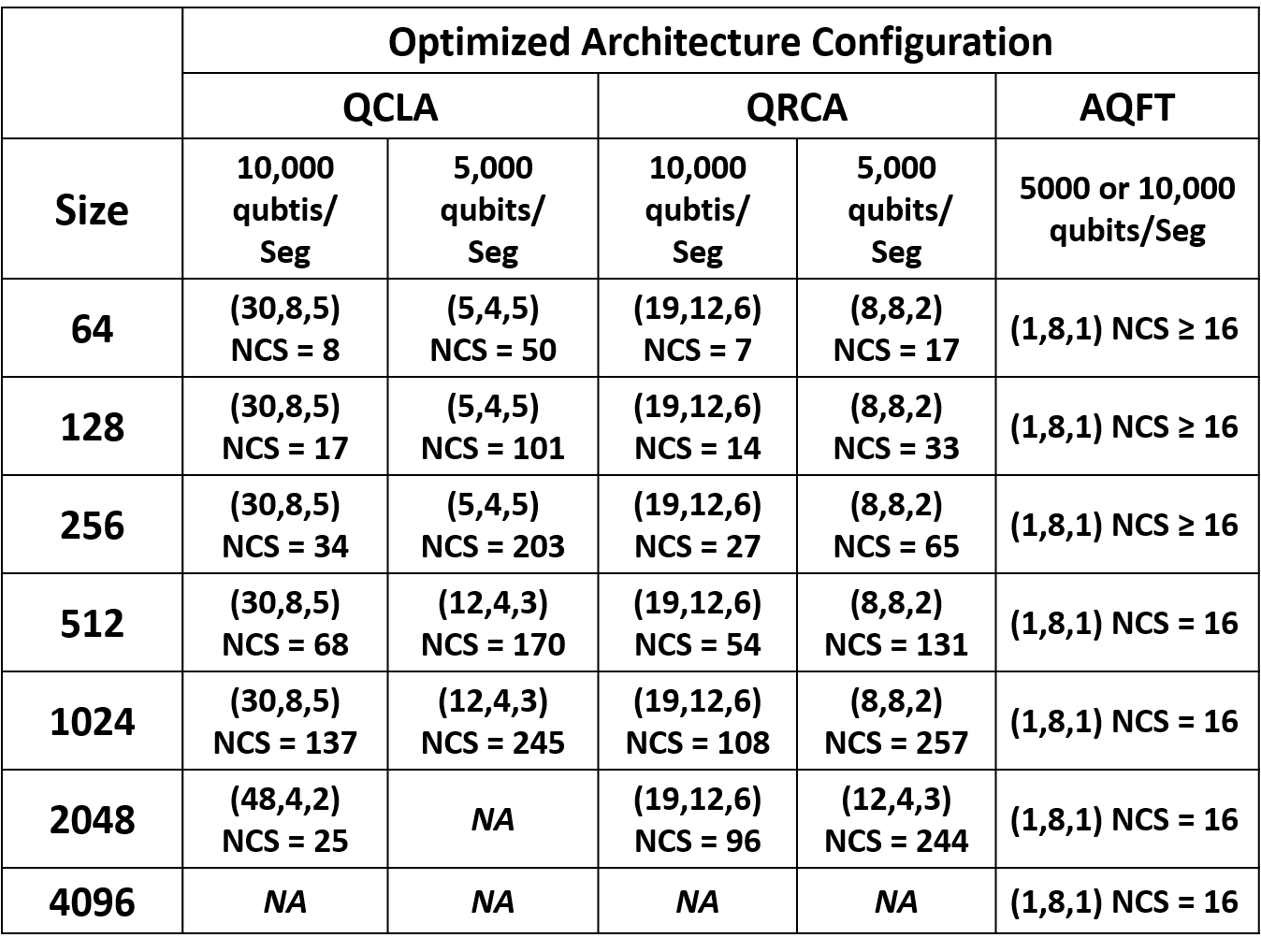}}
\end{table}


\subsection{Design Optimization by Tuning Device Parameters}
\subsubsection{Reducing the execution time}

\begin{figure*}[tb]
	\begin{center}
		\includegraphics[width=1\textwidth]{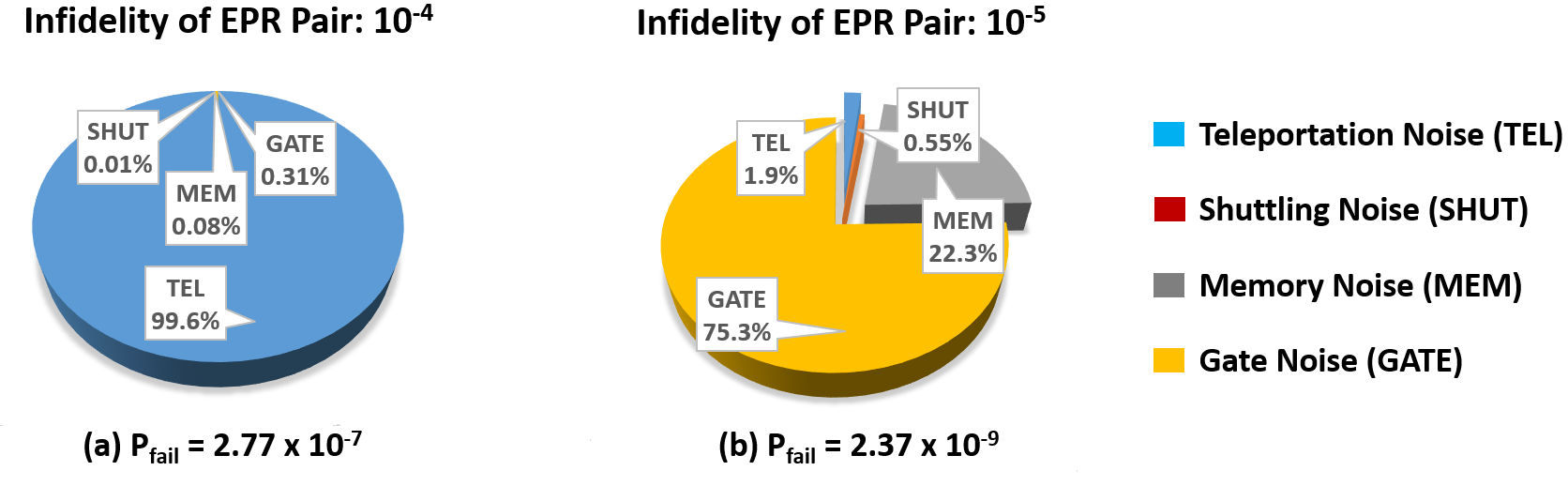}
	\end{center} 
	\caption{The breakdown of and reduction in the failure probability of 2,048-bit QCLA for the configuration [30,8,5], NCS = 273 with $T_{shutt} = 1\mu s$ and the baseline physical measurement and multi-qubit gate times reduced by $90\%$. The original failure probability shown in (a) is reduced from $2.77 \times 10^{-7}$ to $2.37 \times 10^{-9}$ in (b) only by decreasing the infidelity of the EPR pair for cross-Segment communication.} 
	\label{fig:QCLA_Fail}		
\end{figure*}

The largest integer factorized using a classical computers to date is 768 bits, consuming thirty months on a cluster of several hundred processors for the factorization~\cite{kleinjung2010factorization}.  Kleinjung et al. estimated that factorizing a 1,024-bit number is 1000 times harder than a 768-bit number and larger integers are unlikely to be factored in the near future. Based on these reasons, we choose performance criterion so that a QC is considered practical if integers larger than 768 bits can be factored in less than five months.

The total execution time of Shor's algorithm is heavily dominated by the modular exponentiation circuit constructed using adders. Efficient implementation of modular exponentiation for 512-, 1,024- and 2,048-bit integers require 1,4 and 16 million calls to an adder circuit respectively~\cite{VanMeterPRA2005}. Given the size limit of our QC, the adders are assumed to be executed sequentially since parallel implementation of adders will require additional qubit resources. The upper bound for the execution time of an adder circuit to complete the modular exponentiation in five months is shown as dotted line in Fig.\ref{label:Pic_Problem_Scale}. If the execution time of an adder falls below this line, an integer of this size can be factored in less than five months.


Figure~\ref{label:Pic_Problem_Scale} shows that a 1,024-bit number can be factored in less than five months with a 1.5-million-qubit QC, as the $T_{exec}$ of the QCLA lies below the black dotted line. However, the $T_{exec}$ of 2,048-bit QRCA and QCLA (58.5s and 270s respectively) are both significantly higher than the required execution time of 0.8s. We first attempt to reduce $T_{exec}$ by increasing our resource budget. With twice as many qubits, the optimized architecture configuration for QCLA changes from (48,4,2), NCS = 25 to (30,8,5), NCS = 273 and shows nearly 100x decrease in $T_{exec}$ (from 270s to 2.76s). This remarkable reduction arises as the additional delays in QCLA execution time due to limited parallel operation is eliminated, and logarithmic depth performance is restored. The QRCA shows only a nominal reduction from 58.5s to 50.1s, as the resource limitation is not the main cause of slow execution time. The logarithmic depth QCLA is the only choice that can meet the threshold execution time criterion, if $T_{exec}$ can be further reduced by factor of 4. Unfortunately, we find that any further increase in resources (and new architecture configuration) fails to gain additional reduction in $T_{exec}$. 

The failure to achieve performance improvement through additional resources highlights the role of device parameters (DPs) in the design space~\cite{Ahsan2015}. The set of DPs which affect the speed of quantum circuit include the latency of  physical operations. We find that when physical gate (and Measurement) times and qubit shuttling latency are reduced 10x, the execution time declines to $0.68s  < 0.8s$, meeting the required adder execution time. The total execution time for 2,048-bit integer factorization in this case can be approximated as the sum of time spent in 16 million calls to the adder ($0.68s \times 16 \times 10^{6} \approx 128$ days) and the $T_{exec}$ of a single run of the 4,096-bit AQFT (less than a day). Therefore the proposed QC design can factor a 2,048-bit number in less than five months. 

\subsubsection{Reducing the failure probability}
In order to ensure that the entire Shor's algorithm is reliably executed, we require the sum of the failure probability of a 2,048-bit modular exponentiation circuit and a 4,096-bit AQFT to be sufficiently low. Table~\ref{label:Tab_Failure_Scale} shows already adequate $P_{fail}$ in range $\sim O(10^{-4})-O(10^{-3})$ for AQFT when baseline value of $1\mu s$ was assigned to the qubit shuttling latency ($T_{shutt}$). As $T_{shutt}$  reduces to $0.1\mu s$ in our current design to lower the execution time of the adder, $P_{fail}$ of 4,096-bit AQFT falls well below $10^{-4}$. Therefore, the goal of reducing the failure probability of the full Shor's algorithm translates into curtailing the failure probability of the modular exponentiation circuit. This in turn requires the $P_{fail}$ of each adder call to be less than a certain threshold value so that overall failure probability is reduced far enough to meet the design criterion. The 2,048-bit integer factorization consumes 16 million calls to the adder and therefore we require $P_{fail} << 6.67 \times 10^{-8}$ for each adder execution. The $P_{fail}$ for the design optimized for $T_{exec}$ [(30,8,5), NCS = 273] is $2.77 \times 10^{-7}$.     
    
In order to lower the failure probability, we can either add one more layer of encoding or reduce the noise level in the physical device components. Adding a layer of encoding will require at least 7x increase in qubit resources which enormously expands the scale of integration. In addition, the $T_{exec}$ is also significantly inflated (Table~\ref{label:T1} shows that L2 error correction takes 70x more time than L1 error correction). Therefore, increasing the number of layers of encoding achieves a reduction in $P_{fail}$ at the expense of far greater $T_{exec}$ and resources. On the other hand, reducing the noise level in physical device components can decrease $P_{fail}$ without compromising $T_{exec}$. We exploit the tool supplied breakdown of $P_{fail}$ based on the fidelity of component physical operations to systematically improve the success probability of the factorizing task. 
 
Figure~\ref{fig:QCLA_Fail} shows the breakdown of failure probability for 2,048-bit QCLA, where over $99\%$ of the failure originates from Teleportation Noise. The device parameter which directly affects the Teleportation Noise is the infidelity of EPR pairs for cross-Segment communication. By reducing the infidelity from $10^{-4}$ to $10^{-5}$, we gain more than a 100x reduction in the failure probability as shown in Fig.\ref{fig:QCLA_Fail}(b). A further decrease in $P_{fail}$ can be obtained by tuning DPs affecting the Gate and Memory Noise.    
  
In conclusion, we showed that we can lower the failure probability of 2,048-bit QCLA circuit to $2.37 \times 10^{-9}$ ($<< 6.67 \times 10^{-8}$) by tuning the DPs. This gives an overall failure probability of modular exponentiation of about $3.8\%$. The $P_{fail}$ of the 4,096-bit AQFT is negligible compared to this value and the overall failure probability does not exceed $4\%$. Hence we have shown that the optimized adder architecture with the appropriately tuned DPs can be used to construct reliable QC to execute 2,048-bit Shor's algorithm. 

\section{Tool enhancements and extensions} \label{Extensibility}
A .

\section{Comparison with the related architecture work} \label{RelatedWork}
A generation of quantum architectures for large monolithic ion traps had been analyzed using area as a metric for resource utilization~\cite{13,19,21}. Unfortunately, the sizable trap chip envisioned in these studies is difficult to fabricate due to limitations of fabrication technology~\cite{LimitPlanarTrap}. These constraints force us to adopt a modular quantum architecture and a resource metric, both of which are largely decoupled from the trap size.  In this paper we analyzed a multi-core architecture (MUSIQC), which invests in communication qubits to generate EPR pairs in order to connect variable size ion traps (Segments). We used qubits per trap (qubits/Segment) and total qubits in the computer as the resource metrics in our simulations. We find that the MUSIQC architecture directly translates scalability cost into the qubits required for computation and communication regardless of physical size of the trap. By analyzing this modular architecture, we present a practical framework for designing and evaluating future generation of quantum architectures.    

\section{Conclusion} \label{ConSec}
We presented a complete performance simulation toolset capable of designing a resource efficient, scalable QC. Our tool is capable of analyzing the performance metrics of a flexible, reconfigurable computer model, and deepens our insights on the quantum architecture design by providing a comprehensive breakdown of performance metrics and visualization of resource utilization.  Using this tool, we were able to quantify, for the first time, the resource-performance scalability of a proposed architecture, featuring unique properties such as (1) cross-layer optimization, where qubit resources providing L2-level functions are shared throughout the computer, (2) resource-constrained hardware performance, where optimized architectural design for resource allocation is considered as a function of the problem size, and (3) complete visualization of the resource utilization that provides a means to validate the optimality of the performance, (4) over a hardware architecture that provides global connectivity among all the qubits in the system. 

Due to the macro-modeling approach used in our tool, we achieve highly efficient runtime for the performance simulation, which allows us to carry out comprehensive search for an optimized system design under given resource constraints, over a range of architecture configurations and benchmark circuits. Our benchmarks included crucial building blocks of Shor's algorithm, including the approximate quantum Fourier transform and two types of quantum adders. We found that the optimized designs vary across the benchmark applications depending on the types of gates used, the depth and parallelism of circuit structure, and resource budget. By comparing their performance across these benchmark circuits, we present a concrete quantum computer design capable of executing 2,048-bit Shor's algorithm in less than five months.




\bibliographystyle{ACM-Reference-Format-Journals}
\bibliography{JETC_Bib}

\end{document}